\documentclass{aa}
\usepackage[colorlinks=true,linkcolor=blue, citecolor=blue, filecolor=blue, urlcolor=blue]{hyperref}
\usepackage[varg]{txfonts}
\usepackage{amsmath}
\usepackage{algorithmic}
\usepackage{multirow}
\usepackage[lined,ruled,linesnumbered]{algorithm2e}
\usepackage{graphicx}
\usepackage{tikz}
\usetikzlibrary{shapes.geometric,shapes.arrows,decorations.pathmorphing,
decorations.markings,decorations.pathreplacing}
\usetikzlibrary{matrix,chains,scopes,positioning,arrows,fit,spy,calligraphy}  

\begin{document}

\title{Multi-Step Reconstruction of Radio-Interferometric Images}

\author{S. Wang \and S. Prunet \and S. Mignot \and  A. Ferrari}

\institute{Universit\'e C\^ote d’Azur, Observatoire de la C\^ote d’Azur, CNRS, 06000 Nice, France \email{sunrise.wang@oca.eu}}

\abstract{The advent of large aperture arrays, such as the currently under construction Square Kilometer Array (SKA), allows for observing the universe in the radio-spectrum at unprecedented resolution and sensitivity. However, these telescopes produce data on the scale of exabytes, introducing a slew of hardware and software design challenges.}
{This paper proposes a multi-step image reconstruction method that allows for partitioning visibility data by baseline length. This enables more flexible data distribution and parallelization, aiding in processing radio-astronomical observations within given constraints.}
{The multi-step reconstruction is separated into two-steps, first reconstructing a low-resolution image with only short-baseline visibilities, and then using this image together with the long-baseline visibilities to reconstruct the full-resolution image. The proposed method only operates in the minor-cycle, and can be easily integrated into existing imaging pipelines.}
{We show that our proposed method allows for partitioning visibilities by baseline without introducing significant additional drawbacks, having roughly the same computational cost and producing images of comparable quality to a method in the same framework that processes all baselines simultaneously.}
{}

\maketitle 
\nolinenumbers

\section{Introduction} 
\label{S:intro}
Radio-interferometry allows us to obtain images of the sky in the radio spectrum by using antenna arrays in tandem with aperture synthesis. The upcoming Square Kilometer Array Observatory (SKAO)\footnote{\url{https://www.skao.int/}} is composed of two separate telescopes, one for high frequencies ($350$~MHz to $15.35$~GHz) and one for low frequencies ($50-350$ MHz), that are currently being built respectively in South Africa and Australia. They will, upon completion, have unprecedented resolution and sensitivity, enabled by the 197 dishes in SKA-Mid in South Africa, and the 512 antenna stations in SKA-Low in Australia. 

With the large numbers of antennas comes an equally large amount of data. For SKA-Mid, projections estimate up to $2.375\;\text{TB/s}=205.2\; \text{PB/day}$ from the dishes to the beamformer and correlator engines, and $1.125\;\text{TB/s}=97.2\;\text{TB/day}$ from these to the imaging super computer, the SKA-Mid Science Data Processor~\citep{skamid-spie22}. For SKA-Low, the estimated data transfer is $0.725\;\text{TB/s}\approx62.5\;\text{PB/day}$ from the antennas to the correlator, and $0.29\;\text{TB/s}\approx25\;\text{PB/day}$ from the correlator to the SKA-Low Science Data Processor~\citep{skalow-spie22}. Such amounts of data naturally leads to hardware and software design challenges. These include: transferring data between nodes turns out to be very expensive both in terms of computational time and energy; long term data storage proves impossible given the cost; memory usage per Science Data Processor node is a potential concern. 

With these complications comes the need to efficiently partition both the data and the workload. This is typically performed along the frequency and time domains, with reconstruction being performed independently for each partition.  Approaches that separate the image into facets for direction-dependent calibration, such as ~\citet{cornwell1992radio,van2016lofar,tasse2018faceting}, also enable partitioning by the spatial image domain. In addition to the above, one can also potentially separate the sample data i.e. visibilities based on the length of their corresponding baselines. However, this is typically not done due to current minor-cycle reconstruction algorithms needing to process all baselines together to achieve full resolution. This limits parallelization flexibility e.g. de/gridding needs to wait for the reconstruction algorithm to finish processing all baselines before restarting, and memory access patterns become complex for large numbers of baselines due to the large number of visibilities needing to be gridded to the same grid.

This paper proposes an image reconstruction method, in the family of algorithms that are based on compressed sensing and convex optimization~\citep{wiaux2009compressed, carrillo2014purify, garsden2015lofar}, that allows for the separation of visibilities by baseline length. It achieves this by performing reconstruction in two steps, each processing only a subset of the total visibilities. The first produces a low-resolution image using only the short baseline visibilities. The second produces the final reconstructed image, using both the long baseline visibilities, as well as the low-resolution image of the first step.

We evaluate our method in the context of the traditional major-minor cycle pipeline, running a separate pipeline for each step. We found it to produce images of comparable quality in similar amounts of computational time compared to a single-step approach that processes all baselines simultaneously.

The remainder of this paper is structured as follows: We provide a brief overview of radio-interferometric imaging as well as a literature review in Sect.~\ref{S:RadioInterferometry}; we describe our method in Sect.~\ref{S:Algorithm}; Sect.~\ref{S:Datasets} describes the datasets used for our experiments; we present and discuss our results in Sect.~\ref{S:Results}; and we provide conclusions and discuss possible avenues for future work in Sect.~\ref{S:ConclusionFuturework}.

\section{Radio-Interferometric Imaging}
\label{S:RadioInterferometry}
Radio-interferometers measure the sky using arrays of antennas i.e. aperture arrays. Baselines produce visibilities, the correlated instrumental response of the electrical field for some given time duration and electro-magnetic frequency. These can be defined using the measurement equation~\citep{smirnov2011revisiting}:
\begin{equation}
\label{Eq:RIME}
\begin{aligned}
& V(u, v) = C_{uv}\iint D_{uv}(l, m) \frac{I(l, m)}{n} e^{-2\pi i(ul + vm + w(n - 1))}\, dl\, dm \\
& n = \sqrt{1 - l^2 - m^2}
\end{aligned}
\end{equation}
where $C$ represents the direction independent effects, such as antenna gain, $D$ denotes the direction dependent effects, such as phase gradients caused by the Earth's ionosphere, $(u, v, w)$ is the difference between antenna coordinates in the frame where $w$ is aligned with the phase center, $(l, m)$ are the spatial angular coordinates on the celestial sphere, which is also the domain of the integral, and $I$ is the incident radiance of the sky emission.

If the $D$ and $e^{-2\pi iw(n-1)}$ terms are ignored, Equation~\ref{Eq:RIME} simplifies to a 2-dimensional Fourier transform, allowing us to retrieve an image of the sky emission through its inversion. This image contains artefacts caused both by the partial sampling of the Fourier domain due to the geometry of antenna arrays, and by the omission of the $w$ and $D$ terms. Radio-interferometric imaging aims to correct for these, reconstructing an image that is useable for science.

\begin{figure}[t]
\centering
\begin{tikzpicture}[auto, node distance=2cm, >=latex]
	\tikzstyle{block} = [draw, minimum height=3em, minimum width=6em]
	
    \node [name=visibilities] {};
    \node [block, right of=visibilities] (deggrid) {De/grid};
    \node [block, right of=deggrid, node distance=4cm] (deconv) {Deconvolution};

    \draw [->] (deggrid) -- node[name=u] {$\tilde{\imath}_n$} (deconv);
    \coordinate [below of=u] (tmp);

    \draw [->] (visibilities) -- node {$v$} (deggrid);
    
    \node [block, below of=u] (imup) {Image update};
    
    \draw [->] (deconv) |-  node[right] {$i_{n}$} (imup);
    \draw [->] (imup) -| node[left] {$\hat{\imath}_{n+1}$} (deggrid);

\end{tikzpicture}
\caption{A high-level overview of the radio-interferometric pipeline. A reconstructed image $\hat{\imath}_n$ is compared to the measurements $v$ in the de/gridding step, which outputs the difference in the spatial domain $\tilde{\imath}_n$. This is passed to the reconstruction algorithm which generates the next estimate $\hat{\imath}_{n+1}$ by deconvolving the residual image $\tilde{\imath}_n$ and adding it to~$\hat{i}_n$.}
\label{Fig:RIPipelineOverview}
\end{figure}
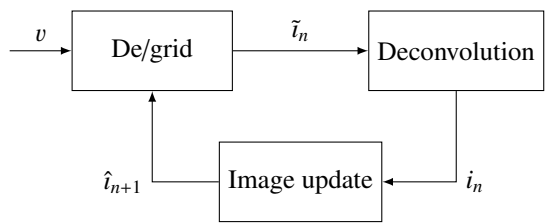

The general approach used by radio-interferometric algorithms is iterative, and is illustrated in Fig.~\ref{Fig:RIPipelineOverview}. The current image estimate $\hat{\imath}_n$ is evaluated against the measurements $v$ in the de/gridding step D/G, which may also correct for the $w$ and $D$ terms. This step produces the difference between $v$ and $\hat{\imath}_n$ in image space, termed $\tilde{\imath}_n$. It can be expressed as
\begin{equation}
\tilde{\imath}_n= F^\dagger G^\dagger(v - GF\hat{\imath}_n)
\end{equation}
where $F$ and $F^\dagger$ are the Fast Fourier Transform (FFT) and its inverse respectively, $G$ is a degridding operator, responsible for resampling the regular gridded visibilities to their original irregular positions, and $G^\dagger$ is the gridding operator, responsible for resampling the irregular visibilities to regular gridded positions.
The correction of the $w$ and $D$ terms typically occurs before $F^\dagger$. There are various approaches to this, such as: projecting the visibilities onto some plane where $w$ and $D$ are zero~\citep{cornwell2008noncoplanar, bhatnagar2008correcting, van2018image}; discretizing in the w-plane, which can be seen in methods such as w-stacking~\citep{offringa2014wsclean} and its improvement~\citep{ye2022high}; discretizing in the time domain, such as with the snapshots method~\citep{ord2010interferometric}; discretizing in the spatial domain with facet-based approaches~\citep{cornwell1992radio, tasse2018faceting}; or some hybrid of the above, such as between w-projection and snapshots~\citep{cornwell2012wide} and w-projection and w-stacking~\citep{pratley2019fast}.

The residual image $\tilde{\imath}_n$ is then sent to the deconvolution algorithm which removes the partial sampling artefacts. There are a plethora of methods that aim to achieve this, with some examples being CLEAN and its variants~\citep{hogbom1974aperture, cornwell2008multiscale, rau2011multi} that deconvolve $\tilde{\imath}_n$ in a greedy non-linear manner much akin to matching pursuit~\citep{Mallat1993}, and convex optimization methods based on Compressive Sensing~\citep{wiaux2009compressed, carrillo2014purify, dabbech2015moresane}. There has also been work done on progressively reconstructing the final image based on sub-sets of visibilities, such as the work of of~\citet{Cai_Pratley_McEwen_2019}, which shares similar general ideas to our approach. It differs in that the partitioning is by time rather than baseline, with their method still requiring all baselines to be present within a sub-set of visibilities. After deconvolution, the image estimate is updated by adding to it the deconvolved residual $i_n$:
\begin{equation}
\hat{\imath}_{n+1} = \hat{\imath}_n + i_n
\end{equation}
The deconvolution step is iterative, ergo the imaging pipeline has a nested loop structure, which is often referred to as the major-minor loop structure, with the loop shown in Fig.~\ref{Fig:RIPipelineOverview} being the major loop, and the deconvolution being the minor.

De/gridding is often the bottleneck of the imaging pipeline due to the sheer number of visibilities needed to be processed, with the work of~\citet{tasse2018faceting} demonstrating that it can reach up to 94\% of the total processing time for a fully serial implementation. Due to this, there is much work that looks to expedite this stage through parallelization. 

This can be done in a coarse or fine grained manner. Fine-grained approaches aim to parallelize on the local machine at a fine-scale, such as per visibility or grid cell. There has been substantial amounts of work done in this area, both for the CPU, such as in~\citet{barnett2019parallel}, as well as the GPU, such as in~\citet{romein2012efficient, merry2016faster, veenboer2017}. Our method does not deal with parallelization on this scale, and focuses primarily on facilitating distribution and parallelization on coarser scales.

Coarse scale parallelization looks to distribute de/gridding across multiple nodes within a cluster. A typical method for this is to distribute according to frequency channels and time. More recently, there has also been work that looks to distribute the gridding according to facets~\citep{monnier2022multi} as well as sections of the uv-plane~\citep{onose2016scalable}. These different distribution strategies can also be combined, for example in the work of~\citet{gheller2023high} where visibilities are separated by time and the v-axis.

An issue with coarse-scaled parallelization strategies that partition in the uv-plane is that the gridded visibilities need to be gathered in a central location, as reconstruction algorithms typically assume that all baselines are processed together. Our proposed method enables more flexibility in this case, as it does not require all baselines to be handled simultaneously. Additionally, although not explored in this paper, it also allows for potential pipeline parallelism, which we elaborate on in Sect.~\ref{SS:FutureWork}.

\section{Proposed algorithm} 
\label{S:Algorithm}

Our proposed algorithm operates within the traditional radio-interferometric imaging framework, specifically replacing the deconvolution step. It performs imaging in two steps, each operating only on a subset of the visibility data separated in the uv-plane. These subsets, termed $\text{V}_\mathcal{L}$ and $\text{V}_\mathcal{H}$ are partitioned according to the domains $\mathcal{L}$ and $\mathcal{H}$, which cover regions representing short and long baselines respectively. 

Each step performs the entire imaging pipeline and produces an image. The first produces a low-frequency image $\hat{\imath}_\mathcal{L}$ from $\text{V}_\mathcal{L}$, and the second produces the fully reconstructed image $\hat{\imath}$ using both $\text{V}_\mathcal{H}$ and $\hat{\imath}_\mathcal{L}$.

The constraint reconstruction framework we use for our work is based on convex optimization with sparsity regularization, as this has a solid theoretical foundation~\citep{candescs2006, donoho2006compressed, candes2011compressed} and has been extensively studied in the field of radio-interferometric imaging~\citep{wiaux2009compressed, carrillo2014purify, dabbech2015moresane}. Specifically, similar to previous works on reconstruction within the LOFAR pipeline~\citep{jiang2015compressed, garsden2015lofar}, we solve for a sparse series of atoms $\alpha$ of a redundant wavelet dictionary $W$. It is important to note that, although detailed for this specific family of reconstruction algorithms, the proposed approach can be adapted to other reconstruction frameworks.

\subsection{Partitioning Visibilities}
\label{SS:Partitioning}

\begin{figure}[t]
\centering
\includegraphics[width=200pt]{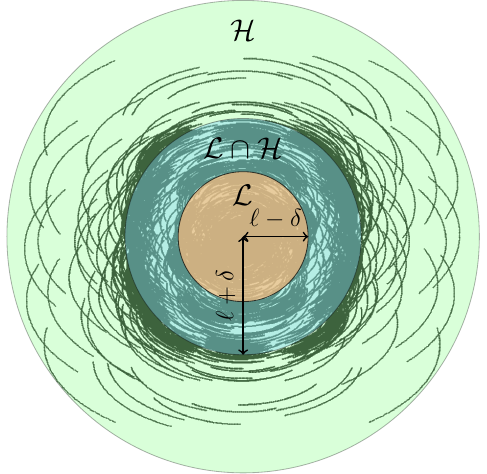}

\caption{The full set of visibilities is partitioned according to domains $\mathcal{L}$ (orange), which contains the short baseline visibilities, and $\mathcal{H}$ (green), which contains the long baseline visibilities. These are not mutually exclusive, but rather have an overlap region, which is denoted in cyan, and is defined using $\ell$, which is the middle of $\mathcal{L} \cap \mathcal{H}$, and $\delta$, the halfwidth of $\mathcal{L} \cap \mathcal{H}$.}
\label{Fig:Partitioning}
\end{figure}

We partition the visibilities into subsets $\text{V}_\mathcal{L}$ and $\text{V}_\mathcal{H}$, based on if the visibilities fall under the domains $\mathcal{L}$ or $\mathcal{H}$ respectively. Figure~\ref{Fig:Partitioning} illustrates this, where visibilities in the orange region are part of $\text{V}_\mathcal{L}$, and visibilities in the green belong to $\text{V}_\mathcal{H}$. It can be seen that $\mathcal{L} \cap \mathcal{H} \neq \emptyset$, and is denoted by the cyan region. This is to alleviate cases where gridding introduces additional spatial frequencies not in the respective domain, caused by the visibilities being interpolated onto the grid through convolution using a kernel with a non-zero support size. The visibilities in $\mathcal{L} \cap \mathcal{H}$ are weighted so that although duplicated, their contribution sums to 1. This is done using filters which are discussed in Sect.~\ref{SS:FullRes}.

The dataset is partitioned using the variables $\ell$ and $\delta$, where $\delta$ defines the halfwidth size of $\mathcal{L} \cap \mathcal{H}$, and $\ell$ defines the center of $\mathcal{L} \cap \mathcal{H}$. These are also shown in Fig.~\ref{Fig:Partitioning}. These values are typically defined in units wavelength, however, we opt to use pixel distances in our paper.

\subsection{Low-Resolution Reconstruction}
\label{SS:LowRes}
The first step involves reconstructing a low-resolution image from $\text{V}_\mathcal{L}$. To do this, for every major-cycle iteration $n$, for a total of $N$ cycles, we solve for the unconstrained problem:
\begin{equation}
\label{Eq:LowRes}
\begin{aligned}
\alpha_{\mathcal{L}_n} &= \arg\min_{\alpha} \|\tilde{\imath}_{\mathcal{L}_n} - H_\mathcal{L} W\alpha\|_2^2 +  \lambda_{\mathcal{L}_n}\|\alpha \|_1\\
i_{\mathcal{L}_n} &= W \alpha_{\mathcal{L}_n}
\end{aligned}
\end{equation}
where $\tilde{\imath}_{\mathcal{L}_n}$ is the current residual image (i.e. between $\text{V}_\mathcal{L}$ and $\hat{\imath}_{\mathcal{L}_{n-1}}$, computed in the $n$th major-cycle), $H_\mathcal{L}$ is the operator detailing the convolution by the PSF associated to  $\text{V}_\mathcal{L}$, $\lambda_{\mathcal{L}_n}$ is the regularization parameter associated to the current major-cycle, and $i_{\mathcal{L}_n}$ is the deconvolved residual of $\tilde{\imath}_{\mathcal{L}_n}$. The final image after $N$ major cycles is then $\hat{\imath}_\mathcal{L} = \tilde{\imath}_{\mathcal{L}_{N + 1}} + \sum^N_{n=1} i_{\mathcal{L}_n}$.

An undesirable side effect when solving for Equation~\ref{Eq:LowRes} is that $\hat{\imath}_\mathcal{L}$ can contain frequency information that is not associated with $\mathcal{L}$ due to $W$ not being constrained to $\mathcal{L}$, which interferes with the reconstruction of the full resolution image. We handle this in the full-resolution reconstruction step using filtering, which we discuss in Sect.~\ref{SS:FullRes}.

\subsection{Full-Resolution Reconstruction}
\label{SS:FullRes}
The second step reconstructs the full resolution image using both $\text{V}_\mathcal{H}$ and $\hat{\imath}_\mathcal{L}$. We formulate the reconstruction problem using two data fidelity terms, for the high and low frequencies respectively, in addition to an $L^1$ regularization term. For every major-cycle iteration $n$, over $N$ total cycles, we solve for:
\begin{equation}
\label{Eq:FullRes}
\begin{aligned}
&\begin{aligned}
\alpha_{\mathcal{H}_n} = &\arg\min_{\alpha} \|G_\mathcal{H} (\tilde{\imath}_{\mathcal{H}_n} - H_\mathcal{H} W\alpha)\|_2^2 + \| G_\mathcal{L} (l_{\mathcal{L}_n} - W\alpha)\|_2^2 \\ &+\lambda_{\mathcal{H}_n}\|\alpha \|_1
\end{aligned}\\
&l_{\mathcal{L}_n} = \hat{\imath}_\mathcal{L} - \sum^{n-1}_{j=1} i_j,\; i_n = W\alpha_{\mathcal{H}_n}
\end{aligned}
\end{equation}
where $\tilde{\imath}_{\mathcal{H}_n}$ is the current residual image between $\text{V}_\mathcal{H}$ and $\hat{\imath}_{n-1}$, $H_\mathcal{H}$ the operator detailing convolution by the PSF associated to $\text{V}_\mathcal{H}$, $\lambda_{\mathcal{H}_n}$ the regularization parameter for the $n$th major-cycle of the full-resolution step. Operator $G_\mathcal{L}$ denotes a low-pass linear filter that discards frequencies not in $\mathcal{L}$ and $G_\mathcal{H}$ is a high-pass linear filter that discards frequencies not in $\mathcal{H}$. Finally, $i_n$ is the deconvolved residual image using both $\tilde{\imath}_{\mathcal{H}_n}$ and $\hat{\imath}_\mathcal{L}$. The final reconstructed image is computed similarly to the low-resolution step, and is $\hat{\imath} = \tilde{\imath}_{\mathcal{H}_{N + 1}} + \sum^N_{n=1} i_n$. 

In Equation \ref{Eq:FullRes}, the first data fidelity term aims to reconstruct the high frequency components of the residual image from $V_\mathcal{H}$, i.e. $\tilde{\imath}_{\mathcal{H}_n}$ for the $n$th major loop. The second data fidelity term guarantees that the low frequency components of the reconstructed image match the reconstruction obtained using  $V_\mathcal{L}$ in the first low-resolution resconstruction step: for the $n$th major loop the low-pass filtered reconstructed image must match the residual between $\hat{\imath}_\mathcal{L}$ and the low-frequency components of the reconstructions at the previous major loops iterations: $G_\mathcal{L}\sum^{n-1}_{j=1} i_j$.
As mentioned in the previous section, $\hat{\imath}_\mathcal{L}$ may contain spatial frequencies not in $\mathcal{L}$ due to $W$ not being limited to $\mathcal{L}$, which may bias the reconstruction of $\hat{\imath}$ towards these. These components are removed by applying $G_\mathcal{L}$ also to $\hat{\imath}_\mathcal{L}$ as in Equation \ref{Eq:FullRes}. 

The gains of the two filters have to be properly normalized, in particular in $\mathcal{L} \cap \mathcal{H}$ to account for visibilities being duplicated. Assume that the two filters $G_\mathcal{L}$ and $G_\mathcal{H}$ have circularly symmetric frequency responses denoted as $g_\mathcal{L}(r)$ and  $g_\mathcal{H}(r)$. Consider that $l_{\mathcal{L}_n}$ contains some reconstruction noise with variance $\eta^2$, and $\tilde{\imath}_{\mathcal{H}_n}$ has noise with variance $\sigma^2$. This suggests to define the two filters as:
\begin{align}
& r> \ell + \delta: \quad |g_\mathcal{H}(r)|^2 = 1/\sigma^2,\; g_\mathcal{L}(u) = 0\\
&r< \ell - \delta: \quad g_\mathcal{H}(r) = 0,\; |g_\mathcal{L}(r)|^2 = 1/\eta^2\\
&\ell - \delta <r< \ell + \delta: \quad \sigma^2|g_\mathcal{H}(r)|^2 + \eta^2|g_\mathcal{L}(r)|^2=1
\end{align}
These constraints ensure normalization of the variance noise across the three frequency domains $\mathcal{L}-\mathcal{H}$, $\mathcal{H}-\mathcal{L}$ and $\mathcal{H}\cap\mathcal{L}$

\begin{figure}[t]
\centering
\includegraphics[width = \columnwidth]{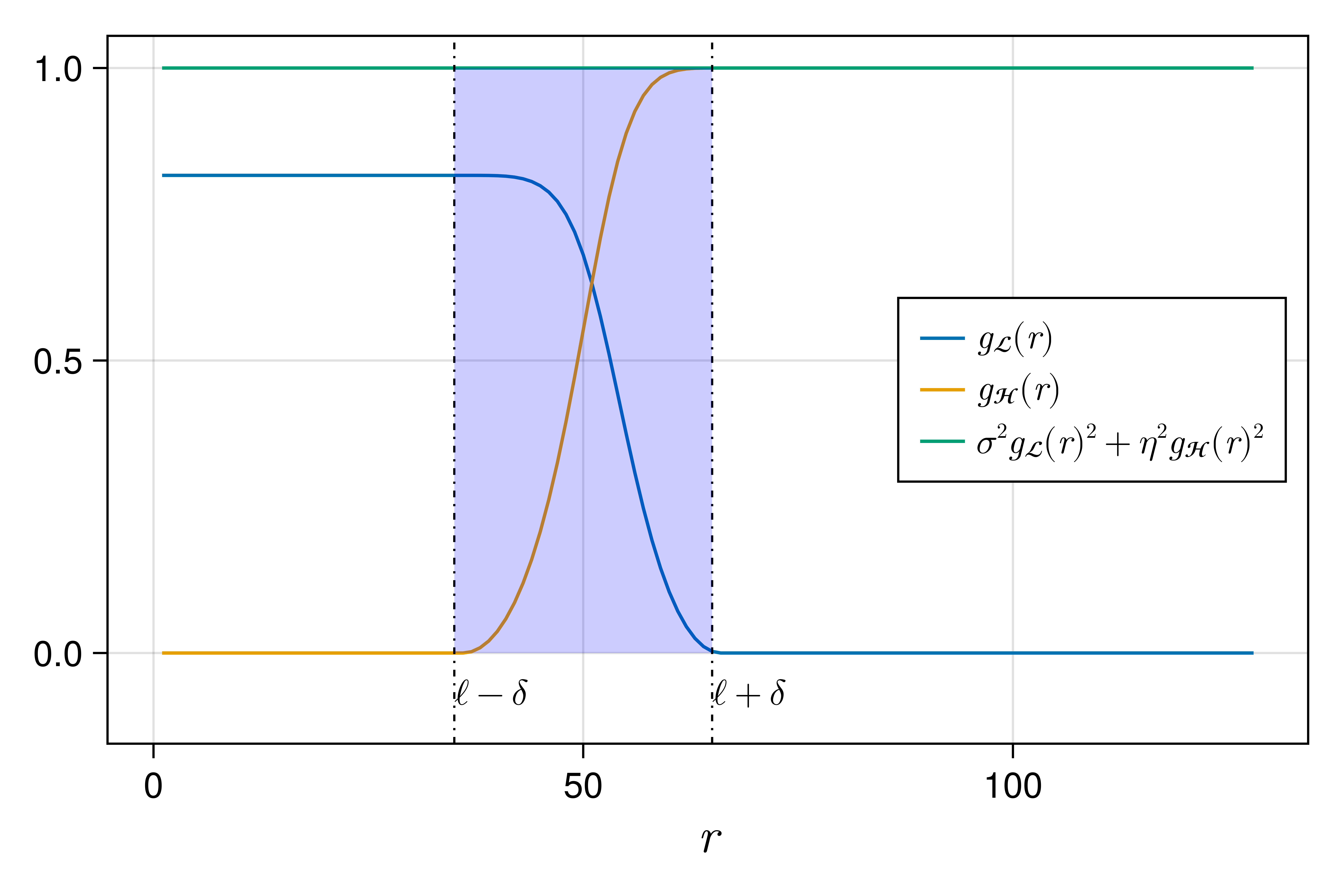}
\caption{Radial frequency response of $G_\mathcal{H}$ and $G_\mathcal{L}$ for $\sigma^2=1$ and $\eta^2=1.2$.}
\label{Fig:FilterResponse}
\end{figure}

We propose herein to define the frequency response of the filters for $\ell - \delta <r< \ell + \delta$ as:
\begin{align}
&g_\mathcal{L}(r) = \alpha(r)\left(1-\sin\left(\frac{\pi}{2\delta}(r - \ell)\right)\right) \\
&g_\mathcal{H}(r) = \alpha(r)\left(1+\sin\left(\frac{\pi}{2\delta}(r - \ell)\right)\right)
\end{align}
where $\alpha(r)$ is such that $\eta^2g_\mathcal{L}(r)^2 + \sigma^2g_\mathcal{H}(r)^2=1$. An example is given in Fig.~\ref{Fig:FilterResponse}.

\subsection{Optimization Algorithm}
\label{SS:OptimizationAlgorithm}
We opt to use the Fast Iterative Shrinkage-Thresholding Algorithm (FISTA)~\citep{beck09} for our optimization algorithm. FISTA is a fast converging algorithm aimed at optimizing convex problems that comprise of both a smooth term and a term that has an easy to solve proximal operator. Equations~\ref{Eq:LowRes} and~\ref{Eq:FullRes} both fall under this umbrella. 

\begin{algorithm}

\begin{algorithmic}[1]
\STATE Initialize $\beta_p$, $\alpha \gets \beta_p$
\FOR{$k=1\ldots N-1$}
\STATE{$\beta \gets \tau_{\gamma}(\alpha - \mu \nabla f(\alpha)) $}
\STATE{$\alpha \gets \beta + \frac{k-1}{k+2} (\beta - \beta_p)$}
\STATE{$\beta_p \gets \beta$}
\ENDFOR
\RETURN{$\alpha $}
\end{algorithmic}
\caption{FISTA  for $\alpha \rightarrow f(\alpha)+ \gamma \|\alpha\|_1$}
\label{Alg:FISTA}
\end{algorithm}

FISTA operates iteratively and involves computing the gradient of the smooth portion and the proximal operator of the non-smooth, which for the $L^1$-norm is the soft-thresholding operator. It then computes the candidate solution for the next iteration using a gradient step-size $\mu$, a soft-thresholding step-size, and a momentum term to allow for faster convergence. We describe FISTA in Algorithm~\ref{Alg:FISTA}.

FISTA requires the gradients of the objective function for each iteration. These are, for the $n$th major-cycle:
\begin{align}
&\nabla f(\alpha)_{\mathcal{L}_n}=2W^\dagger H^\dagger_\mathcal{L}(H_\mathcal{L}W\alpha - \tilde{\imath}_{\mathcal{L}_n}) \label{Eq:GradL}\\
&\begin{aligned}
\nabla f(\alpha)_{\mathcal{H}_n}=2W^\dagger &\left(H_\mathcal{H}^\dagger G_\mathcal{H}^\dagger G_\mathcal{H}( H_\mathcal{H}W\alpha - \tilde{\imath}_{\mathcal{H}_n}) \right. \\
&\left. + G_\mathcal{L}^\dagger G_\mathcal{L}(W\alpha  - l_{\mathcal{L}_n})\right)
\label{Eq:GradH}
\end{aligned}
\end{align}
for Equations~\ref{Eq:LowRes} and~\ref{Eq:FullRes} respectively. The gradient step-size is set as $\mu = \dfrac{1}{\vartheta}$, where $\vartheta$ is the Lipschitz constant of $\nabla f(\alpha)$, defined for our problem as
\begin{align}
\vartheta_{\mathcal{L}} &= 2\lambda_\text{max}\left(W^\dagger H_\mathcal{L}^\dagger H_\mathcal{L} W\right)\\
\vartheta_{\mathcal{H}} &= 2\lambda_\text{max}\left(W^\dagger  (H_\mathcal{H}^\dagger G_\mathcal{H}^\dagger G_\mathcal{H}H_\mathcal{H} +  G_\mathcal{L}^\dagger G_\mathcal{L})W\right)
\end{align}
for Equations~\ref{Eq:LowRes} and~\ref{Eq:FullRes} respectively, where $\lambda_\text{max}$ is the largest eigenvalue. 

In the case where $W$ consists of the concatenation of $M$ orthogonal decompositions, as in \citep{onose2016scalable}, and the convolutions are circular:
\begin{align}
\vartheta_\mathcal{L} &= 2M\max \{|\hat{H}_{\mathcal{L}}|^2\}\\
\vartheta_\mathcal{H} &= 2M\max \{|\hat{G}_\mathcal{H}\odot\hat{H}_\mathcal{H}|^2 +  |\hat{G}_\mathcal{L}|^2\}
\end{align}
where $\hat{A}$ denotes the DFT of the PSF associated to $A$. In other cases, such as in \citet{starck2007undecimated}, $\lambda_\text{max}$ can be obtained using an algorithm such as power iteration.

Finally, there is the question of how to approximate $\sigma^2$ and $\eta^2$, and how to select the regularization parameters $\lambda_\mathcal{L}$ and $\lambda_\mathcal{H}$. The former will be discussed in Sect.~\ref{SS:VisReconVariance}, whereas the latter in Sect.~\ref{AS:Lambda}.

\subsection{Implementation}
We implement our method in Julia~\citep{bezanson2017julia} and integrate it into the RASCIL~\citep{cornwell_rascil} framework under the clean algorithm name \texttt{mstep}. Our implementation and results, as well as instructions on how to use our code will be made available upon publication on our on-line repository.

\section{Simulated Datasets}
\label{S:Datasets}
\subsection{Full Datasets}
\begin{figure}[t]
\centering
\begin{tikzpicture}
  \node(f1) at (0, 0){
    \includegraphics[width=115pt]{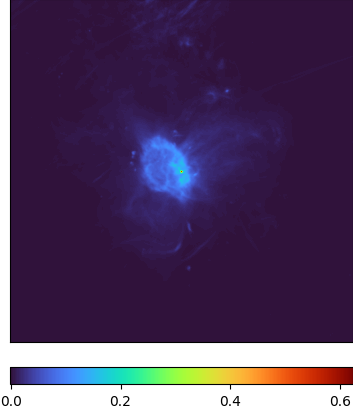}
  };
  \node(f2)[right=1pt] at (f1.east){
    \includegraphics[width=115pt]{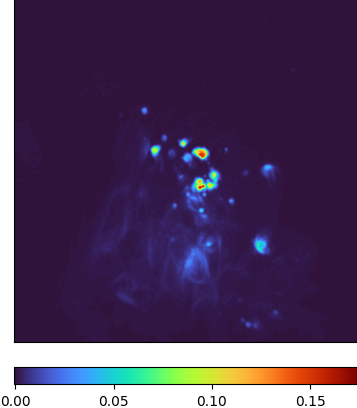}
  };
  \node(f3)[below=1pt] at (f1.south){
    \includegraphics[width=115pt]{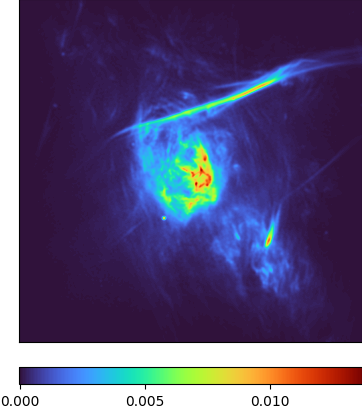}
  };
  \node(f4)[below=1pt] at (f2.south){
    \includegraphics[width=115pt]{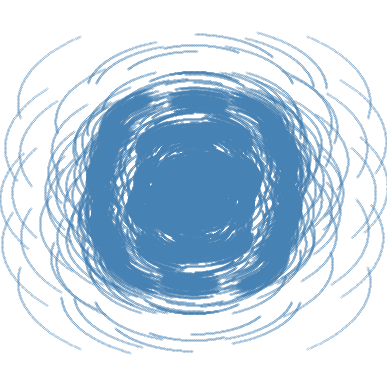}
  };

  \node[text centered, above=0pt, font=\fontsize{3}{0}] at (f1.north){Sgr A};
  \node[text centered, above=0pt, font=\fontsize{3}{0}] at (f2.north){Sgr B2};
  \node[text centered, below=0pt, font=\fontsize{3}{0}] at (f3.south){Sgr C};
  \node[text centered, below= 15pt, font=\fontsize{3}{0}] at (f4.south){UV Coverage};

\end{tikzpicture}
\caption{Ground truth images of our three datasets, obtained by cutting out and tapering sections of the 1.28 GHz Meerkat galactic center mosaic~\citep{heywood20221}. The images are $512 \times 512$ pixels in size, with a resolution of 1.1'' per pixel. The uv coverage for our simulate datasets is shown on the bottom-right. The same coverage is used for all three datasets.}
\label{Fig:GTImages}
\end{figure}
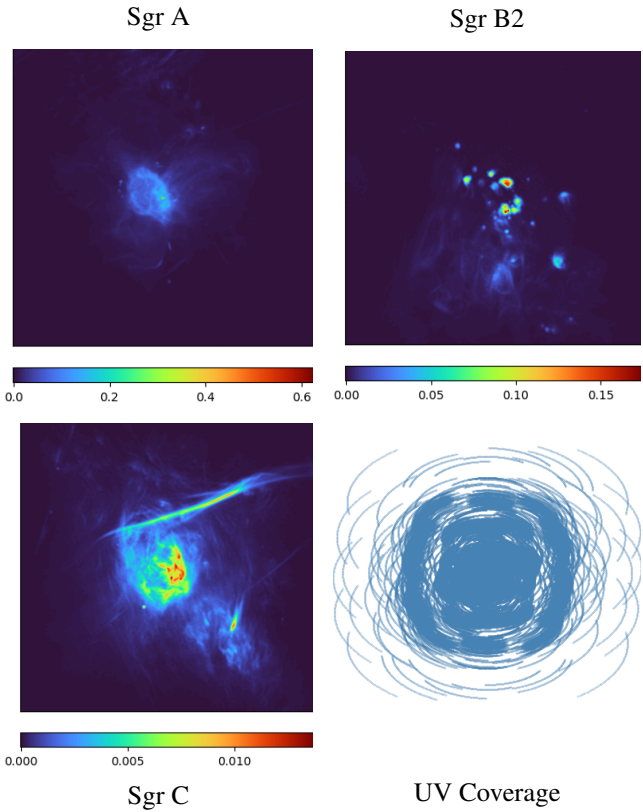

We use three different simulated datasets for our experiments, obtained from $512 \times 512$ pixel tapered cutouts of the 1.28 GHz MeerKAT mosaic~\citep{heywood20221}. These consist of the regions surrounding Sgr A, B2, and C with pixel resolutions of 1.1'', and can be seen along with their uv coverage in Fig.~\ref{Fig:GTImages}. 

We use RASCIL to generate the visibilities. We first generate the visibility positions using a telescope configuration detailing the 64 MeerKAT dishes, assuming the pseudo right ascension and declination of the sources are the same, the observation lasts 4 hours (-2 to 2 hour angles) with visibilities sampled every 120\ s, resulting in 249600 unique positions. We then perform an FFT on our ground-truth images to obtain their respective gridded visibilities, which we degrid to the generated irregular positions with uniform weighting using the improved w-stacking gridder~\citep{ye2022high}. Finally, we model the noise received by antennas by perturbing the visibilities with noise sampled from $\mathcal{N}(0, \sigma/50)$, where $\sigma$ is the standard deviation of the visibilities. A more thorough study of how noise affects our reconstruction method lies outside the scope of this paper.

\subsection{Partitioning Configurations}
\begin{figure}[t]
\centering
\begin{tikzpicture}[spy using outlines={rectangle,black,size=50pt, connect spies}]
\node(f1) at (0, 0){
    \includegraphics[width=220pt]{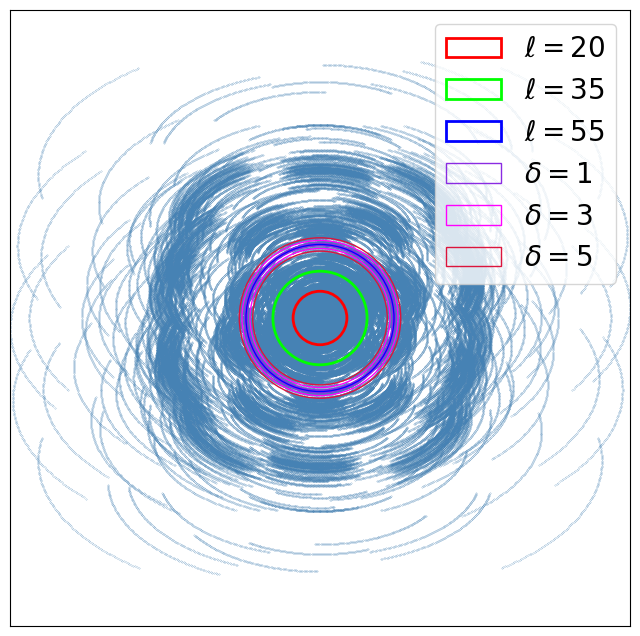}
  };

  \node[xshift=22pt, yshift=10pt](f1 errors spyanchor) at (f1){};
  \spy[magnification=5] on (f1 errors spyanchor) in node [above right=20pt] at (f1.south);
\end{tikzpicture}
\caption{We partition our initial datasets with different values of $\ell$ and $\delta$ to achieve 9 different partitioning configurations per dataset. We show only how $\delta$ for $\ell=55$ in this figure to avoid clutter.}
\label{Fig:DatasetPartitions}
\end{figure}

\begin{table}[ht]
\centering
\begin{tabular}{ c c || c c c}
    & & $\text{V}_\mathcal{L}$ & $\text{V}_\mathcal{H}$ & $\text{V}_{\mathcal{L} \cap \mathcal{H}}$ \\
    \hline\hline
    \multirow{3}{*}{$\ell=20$} 
    &$\delta = 1$ & 106548 & 150641 & 7589\\
    &$\delta = 3$ & 113791 & 160061 & 24252\\
    &$\delta = 5$ & 118941 & 169477 & 38818\\
    \hline
    \multirow{3}{*}{$\ell=35$} 
    &$\delta = 1$ & 141653 & 111758 & 3811\\
    &$\delta = 3$ & 146107 & 115792 & 12299\\
    &$\delta = 5$ & 149271 & 120204 & 19875\\
    \hline
    \multirow{3}{*}{$\ell=55$} 
    &$\delta = 1$ & 167772 & 84034 & 2206\\
    &$\delta = 3$ & 169771 & 85758 & 5929\\
    &$\delta = 5$ & 171508 & 87509 & 9417
\end{tabular}
\caption{Number of visibilities in each region for each partitioning configuration. This applies to all three datasets as they have the same uv coverage.}
\label{Table:Partitionings}
\end{table}

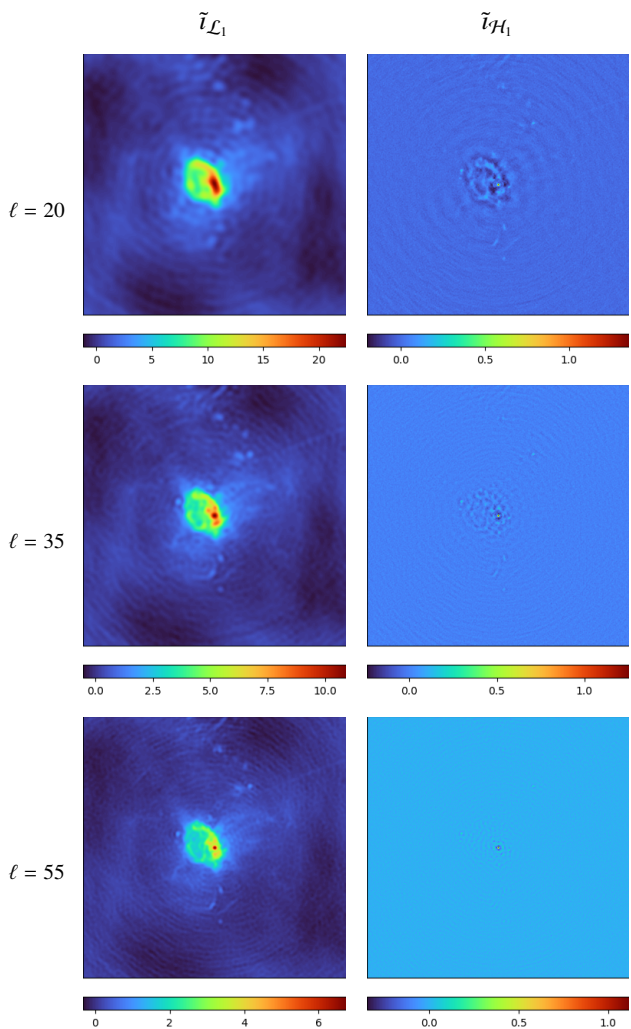
\begin{figure}[t]
\centering
\begin{tikzpicture}
  \node(f1l) at (0, 0){
    \includegraphics[width=98pt]{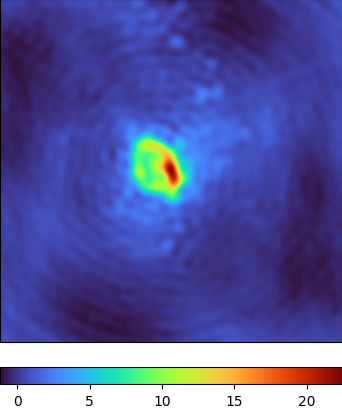}
  };
  \node(f1h)[right=1pt] at (f1l.east){
    \includegraphics[width=98pt]{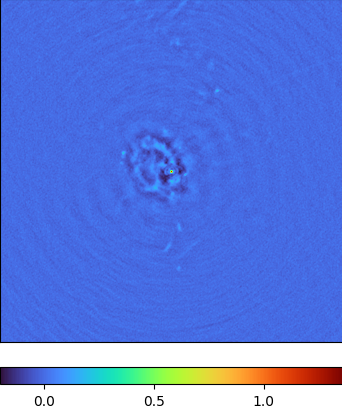}
  };
  \node(f2l)[below=1pt] at (f1l.south){
    \includegraphics[width=98pt]{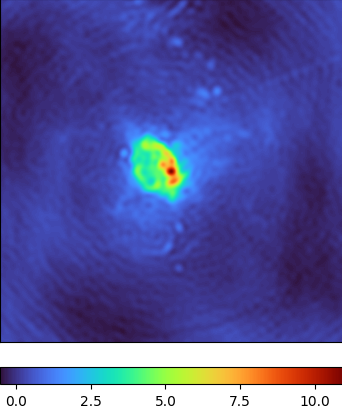}
  };
  \node(f2h)[below=1pt] at (f1h.south){
    \includegraphics[width=98pt]{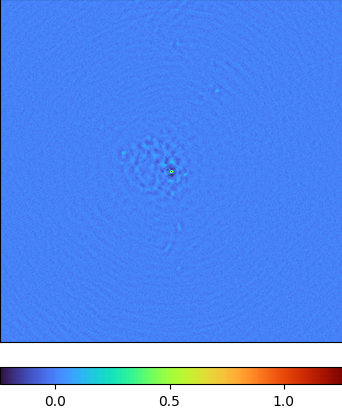}
  };
  \node(f3l)[below=1pt] at (f2l.south){
    \includegraphics[width=98pt]{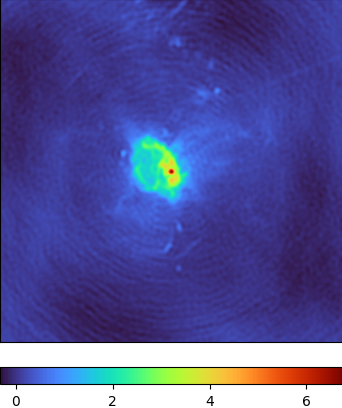}
  };
  \node(f3h)[below=1pt] at (f2h.south){
    \includegraphics[width=98pt]{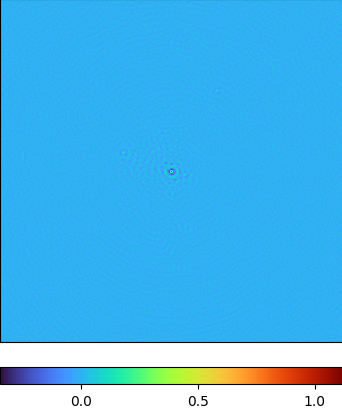}
  };

  \node[text centered, left=0pt, font=\fontsize{8}{0}] at (f1l.west){$\ell = 20$};
  \node[text centered, left=0pt, font=\fontsize{8}{0}] at (f2l.west){$\ell = 35$};
  \node[text centered, left=0pt, font=\fontsize{8}{0}] at (f3l.west){$\ell = 55$};
  \node[text centered, above=0pt, font=\fontsize{10}{0}] at (f1l.north){$\tilde{\imath}_{\mathcal{L}_1}$};
  \node[text centered, above=0pt, font=\fontsize{10}{0}] at (f1h.north){$\tilde{\imath}_{\mathcal{H}_1}$};
\end{tikzpicture}
\caption{Dirty images $\tilde{\imath}_{\mathcal{L}_1}$ and $\tilde{\imath}_{\mathcal{H}_1}$ for different values of $\ell$ for the Sgr A dataset. We set $\delta=5$ for these images.} 
\label{Fig:PartitioningDirties}
\end{figure}

We partition each initial dataset with three different centers of separation $\ell$, each with three different half-width sizes $\delta$, resulting in 9 different partitioning configurations for each. Figure~\ref{Fig:DatasetPartitions} illustrates both $\ell$ and $\delta$ in the context of our datasets, whereas Table~\ref{Table:Partitionings} details the number of visibilities in $\text{V}_\mathcal{L}$ and $\text{V}_\mathcal{H}$ for each configuration. It also details the number of duplicated visibilities, which lie in the region $\mathcal{L} \cap \mathcal{H}$. These numbers are identical for all three of our datasets as they have the same observational parameters. We show an example of how the partition configuration affects the dirty images for the Sgr A dataset in Fig.~\ref{Fig:PartitioningDirties}.

\subsection{Visibility and Reconstruction Variance}
\label{SS:VisReconVariance}
Our proposed method requires knowing the variances $\sigma^2$ and $\eta^2$ of $\tilde{\imath}_\mathcal{H}$ and $\hat{\imath}_\mathcal{L}$ respectively. Ensemble statistics are required to estimate these, and thus they need to be approximated. To derive an appropriate strategy, we first compute estimations using ensemble populations of 50 for both $\text{V}_\mathcal{L}$ and $\text{V}_\mathcal{H}$, each with independently sampled and identical noise properties to the original. 

For $\sigma^2$, we use the average per-pixel variances of the different realizations of $\tilde{\imath}_{\mathcal{H}_1}$ computed from the different realizations of $\text{V}_\mathcal{H}$ using RASCIL. For $\eta^2$, we reconstruct $\hat{\imath}_\mathcal{L}$ over 3 major-cycles for each realization of $\text{V}_\mathcal{L}$, and then set $\eta^2$ to be equal to the average of the per-pixel variance amongst these. For these reconstructions, we use the regularization parameter $\lambda_{\mathcal{L}_n} = 0.05\|\hat{\imath}_{\mathcal{L}_n}\|_2 \times 2^n$ for the $n$th major cycle, as well as the concatenation of the first 8 Daubechies wavelets for our dictionary. These choices are discussed in Sect.~\ref{App:ParamSelection}.

\begin{table}[ht]
\centering
\begin{tabular}{ c c || c c | c c }
    & & $\sigma^2$ & $\hat{\sigma}^2$ & $\eta^2$ & $\hat{\eta}^2$ \\
    \hline\hline
    \multirow{3}{*}{Sgr A} 
    &$\ell = 20$ & 2.8e-04 & 2.2e-04 & 2e-08 & 2.2e-07\\
    &$\ell = 35$ & 3e-04 & 2.3e-04 & 1e-07 & 2.3e-07\\
    &$\ell = 55$ & 3.6e-04 & 2.5e-04 & 2.6e-07 & 2.5e-07\\
    \hline
    \multirow{3}{*}{Sgr B2} 
    &$\ell = 20$ & 2e-05 & 2.2e-05 & 2e-08 & 2.2e-07\\
    &$\ell = 35$ & 2.2e-05 & 6e-05 & 9.5e-09 & 6e-08\\
    &$\ell = 55$ & 2.5e-05 & 3.8e-05 & 2e-08 & 3.8e-08\\
    \hline
    \multirow{3}{*}{Sgr C} 
    &$\ell = 20$ & 4e-06 & 2.8e-06 & 3e-10 & 2.8e-09\\
    &$\ell = 35$ & 4.2e-06 & 2.7e-06 & 1.6e-09 & 2.7e-09\\
    &$\ell = 55$ & 5e-06 & 2.8e-06 & 3.2e-09 & 2.8e-09\\
\end{tabular}
\caption{The estimated values for $\sigma^2$ and $\eta^2$ from populations of 50 realizations, as well as how these compare to our approximations $\hat{\sigma}^2$ and $\hat{\eta}^2$.}
\label{Table:EtaSigmaPartitions}
\end{table}

We approximate $\sigma^2$ with $\hat{\sigma}^2$ by first estimating the variance of pixels in $\tilde{\imath}_{\mathcal{H}_1}$ within a $5 \times 5$ sliding window, and then averaging these values. Approximating $\eta^2$ is more complicated as it is dependent on the details of the reconstruction algorithm in the low-resolution step e.g. an exceedingly large value of $\lambda_\mathcal{L}$ would result in $\eta^2 = 0$ as all realizations of $\hat{\imath}_\mathcal{L}$ will be zero. Rather than derive a strategy for this, we instead apply a constant factor of $\hat{\eta}^2 = \hat{\sigma}^2\,10^{-3}$ for our experiments, as we observed this to be roughly the relationship between the two variables for many cases.

Table~\ref{Table:EtaSigmaPartitions} shows $\eta^2$ and $\sigma^2$ for the different values of $\ell$, as well as our estimated values $\hat{\sigma^2}$ and $\hat{\eta^2}$. We set $\delta=5$ and do not vary it as it does not significantly change the variance. It can be seen that $\hat{\sigma}^2$ is relatively close to $\sigma^2$, diverging by at most $2.75\times$, but in most cases is much closer to the estimated value. Although our approximations for $\eta^2$ are also close in some cases, there are others where they are off by more than $10\times$, with the worst being for the dataset Sgr B2 with $\ell=35$, where $\hat{\eta}^2$ is almost $100\times$ larger. In practice, we found that these variations did not greatly change the quality of the final reconstruction.
However, investigation into better approximation strategies may yield improved convergence speeds.

\section{Simulation results}
\label{S:Results}
We evaluate our proposed multi-step method by first studying how the partitioning configuration affects the final image reconstruction, both in terms of quality and speed. We then compare our method against a method that reconstructs with all baselines in $\mathcal{L} \cup \mathcal{H}$ in a single step, which affords us information on how partitioning visibilities by baseline compares to processing all the visibilities simultaneously without any partitioning.

For our algorithm parameters, we set $\lambda_n$, $\lambda_{\mathcal{L}_n}$ and $\lambda_{\mathcal{H}_n}$ for the $n$th major-cycle to $0.01\|\tilde{\imath}_n\|_2 \times 2^n$, $0.05\|\tilde{\imath}_{\mathcal{L}_n}\|_2 \times 2^n$, and $0.05\|\tilde{\imath}_{\mathcal{H}_n}\|_2 \times 2^n$ respectively, and for $W$, we use a concatenation of the first 8 Daubechies wavelets. We discuss our motivations for these in Sect.~\ref{App:ParamSelection}. Finally, we use a static value of 100 FISTA iterations as the stopping condition for both the single-step full-resolution method, as well for both steps in our multi-step method. 

\subsection{Partition Configuration effect on Reconstruction Accuracy}
\label{SS:MStepAccuracy}
\begin{table}[ht]
\centering
\begin{tabular}{ c c || c | c | c}
    & & Sgr A & Sgr B2 & Sgr C \\
    \hline\hline
    \multirow{3}{*}{$\ell = 20$} 
    &$\delta = 1$ & 22.6 (0.99) & 20.9 (0.92) & 19 (0.97)\\
    &$\delta = 3$ & 22.9 (1) & 21.2 (0.93) & 19.3 (0.98)\\
    &$\delta = 5$ & 21.9 (0.96) & 21.2 (0.93) & 18.5 (0.94)\\
    \hline
    \multirow{3}{*}{$\ell = 35$} 
    &$\delta = 1$ & 22.5 (0.98) & 21.6 (0.95) & 19.6 (1)\\
    &$\delta = 3$ & 22.1 (0.97) & 21.6 (0.95) & 19.6 (1)\\
    &$\delta = 5$ & 21.6 (0.94) & 21.5 (0.94) & 19.2 (0.98)\\
    \hline
    \multirow{3}{*}{$\ell = 55$} 
    &$\delta = 1$ & 21.2 (0.93) & 21 (0.92) & 18.6 (0.95)\\
    &$\delta = 3$ & 21.6 (0.94) & 22.3 (0.98) & 18.8 (0.96)\\
    &$\delta = 5$ & 21.9 (0.96) & 22.8 (1) & 19.2 (0.98)\\
\end{tabular}
\caption{PSNRs (dB) of the various partition configurations for our three datasets with 5 major cycles for each step, with values normalized by the maximum PSNR of each dataset given in parenthesis for easier comparison. It can be seen that there are only minor differences between the different configurations, explainable due to slight differences in ideal reconstruction parameters.}
\label{Table:PartitionPSNRs}
\end{table}

Table~\ref{Table:PartitionPSNRs} compares the peak signal-to-noise ratios (PSNRs) in dB of the final reconstructed images for each partition configuration. These are obtained by running the multi-step reconstruction algorithm for 5 major-cycles, both for the low and full-resolution steps. 

We found that there are only slight differences between the PSNRs when varying $\ell$, without any obvious trend. This leads us to believe that the difference in final image quality is caused mainly by the choice and approximation of $\lambda$, $\sigma^2$, and $\eta^2$. This is advantageous as it implies that one can adjust the partition sizes to suit ones needs without drastically impacting the quality of the final reconstructed image. 
We also found the choice of $\delta$ to not greatly impact the quality of the final image. This could be because it only affects a few spatial frequencies, thus, the affect on final image quality is dominated by other factors, such as parameter selection. Practically, it means that one can decrease $\delta$ without greatly impacting the final reconstructed image, reducing the number of duplicate visibilities. However, care needs to be taken as reducing $\delta$ too much may result in frequency spillage, biasing the reconstruction.

\subsection{Reconstruction Speed}
We evaluate two aspects of reconstruction speed. The number of major-cycles required for convergence, and the number of required FISTA iterations. We evaluate the latter solely in the context of the first major-cycle, as this is where the majority of information is reconstructed. 

\begin{figure}[t]
\centering
\begin{tikzpicture}
  \node(low) at (0, 0){
    \includegraphics[width=200pt]{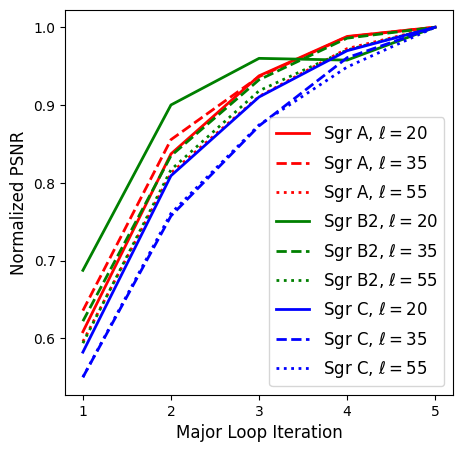}
  };
  \node(high)[below=20pt] at (low.south){
    \includegraphics[width=200pt]{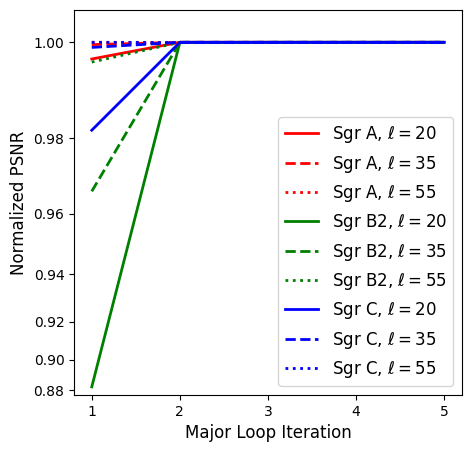}
  };

  \node[text centered, above=0pt, font=\large] at (low.north){Low-Resolution Step using $\text{V}_\mathcal{L}$};
  \node[text centered, above=0pt, font=\large] at (high.north){Full-Resolution Step using $\text{V}_\mathcal{H}$ and $\hat{\imath}_\mathcal{L}$};
\end{tikzpicture}
\caption{Progression of PSNR of reconstructed images for various partitioning configurations across the major cycles for both the low and full resolution reconstruction steps. PSNRs are normalized by the maximum value of each configuration to allow for easier comparison. An exponential scale is used for the bottom image to better illustrate the differences between the different configurations. We fix $\delta=5$ here as we found it not to greatly impact the results.}
\label{Fig:MajCycleConvergence}
\end{figure}

Figure~\ref{Fig:MajCycleConvergence} shows the progression of the PSNRs of the reconstructed images across major cycles. We normalize these by the maximum obtained PSNR for the configuration across the major-cycles for easier comparison. We found that for the low-resolution step, a lower value of $\ell$ often results in a minor increase in convergence speed. This is expected, as a lower $\ell$ corresponds to a simpler image to reconstruct. However, this is counteracted in the second step, where the corresponding configurations have more information and are more challenging to reconstruct. This can also be seen for the Sgr B2 and Sgr C datasets, which have more information in $\tilde{\imath}_{\mathcal{H}_1}$ and thus have worse initial reconstruction quality than the Sgr A dataset.

A surprising result is that the full-resolution step converges after at most only two major cycles. This may be explained by a reduced gridding error for the long baselines, leading to most of the structure being reconstructed quickly. The remaining structure is fine-scale and intermingled with noise, and thus requires very precise values of $\lambda_{\mathcal{H}_n}$ to reconstruct, with a value that is too small detracting from the final reconstructed image, or in our case, a value too large leading to a blank image as the regularization term dominates, resulting in all wavelet coefficients being zero.

\begin{figure}[t]
\centering
\begin{tikzpicture}
  \node(low) at (0, 0){
    \includegraphics[width=200pt]{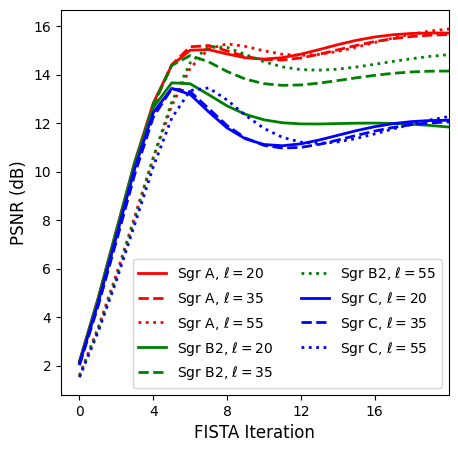}
  };
  \node(high)[below=20pt] at (low.south){
    \includegraphics[width=200pt]{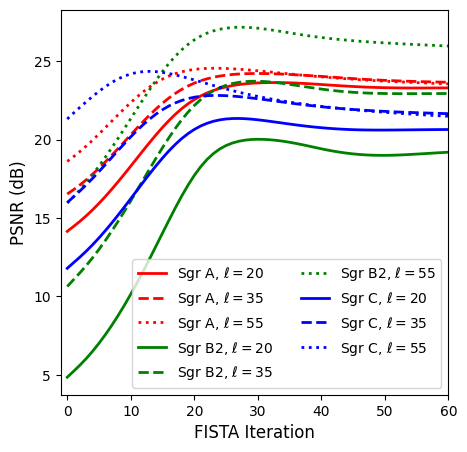}
  };

  \node[text centered, above=0pt, font=\large] at (low.north){Low-Resolution Step with $\tilde{\imath}_{\mathcal{L}_1}$};
  \node[text centered, above=0pt, font=\large] at (high.north){Full-Resolution Step with $\tilde{\imath}_{\mathcal{H}_1}$ and $\hat{\imath}_\mathcal{L}$};
\end{tikzpicture}
\caption{PSNR of the candidate solution of the specified FISTA iteration for the first major cycle of the low and full-resolution steps respectively. The full-resolution images were constructed with $\hat{\imath}_\mathcal{L}$ set as a low-pass filtered version of the ground truth image. We observe slightly faster reconstruction speeds for low values of $\ell$ in the low-reconstruction step, but these configurations are significantly slower in the full-resolution step. It should be noted that although lower values of $\ell$ improve faster for the low-resolution step, they do not necessarily converge at a better result, as illustrated in the dataset Sgr B2.}
\label{Fig:FISTAConvergence}
\end{figure}

Figure~\ref{Fig:FISTAConvergence} shows the PSNR of the candidate solutions across FISTA iterations for both reconstruction steps. We set $\hat{\imath}_\mathcal{L}$ to be a low-pass filtered version of the ground truth for the full-resolution step for these reconstructions.

We observe a slight change in the low-resolution step reconstruction speed when varying $\ell$. On the contrary, we observe a much more significant change in the full-resolution step. This is expected, as the low-resolution reconstruction is simpler than the full and uses less FISTA iterations, thus, changes are less pronounced. Note that for the low-resolution step, although lower values of $\ell$ typically result in faster initial improvement, they do not guarantee that the final image will be of higher quality, such as with the Sgr B2 dataset. However, there are additional influencing factors in these cases, such as differences in ideal parameters, thus it is inconclusive whether the final difference in image quality is due to $\ell$ specifically. It should be noted that despite this, the different configurations all converge to roughly similar PSNR values, as shown in Sect.~\ref{SS:MStepAccuracy}.

Taking into account the reconstruction speeds for both major-cycles and FISTA iterations, the slight change in FISTA iterations is counteracted by the larger number of required major-cycles for the low-resolution step, and vice-versa for the full-resolution step. Thus, we can conclude that the choice of $\ell$ and $\delta$ should not greatly change the reconstruction speed as the gains and losses roughly cancel out.

\subsection{Comparison to Reconstruction without Visibilities Partition}
\begin{figure}[t]
\centering
\begin{tikzpicture}
  \node(a) at (0, 0){
    \includegraphics[width=100pt]{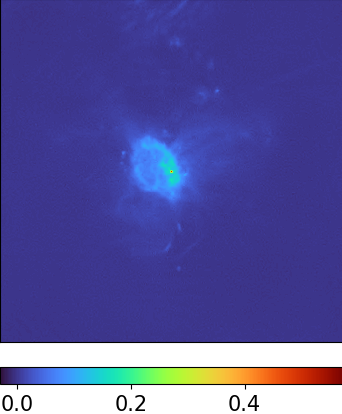}
  };
  \node(b)[below=0pt] at (a.south){
    \includegraphics[width=100pt]{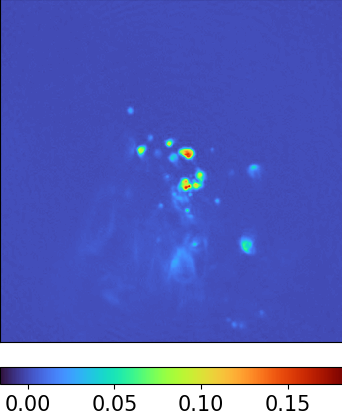}
  };
  \node(c)[below=0pt] at (b.south){
    \includegraphics[width=100pt]{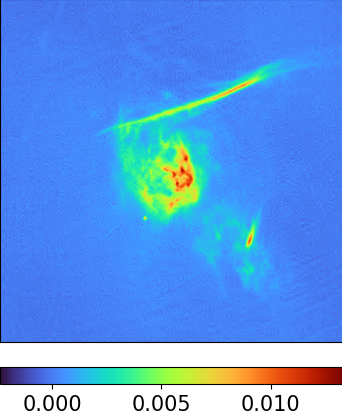}
  };

  \node(am)[right=0pt] at (a.east){
    \includegraphics[width=100pt]{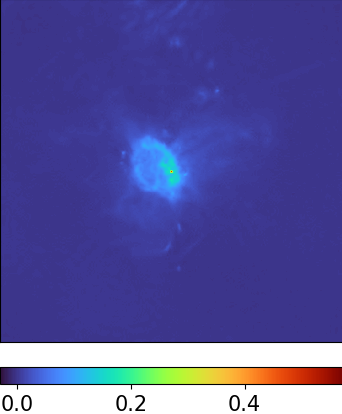}
  };
  \node(bm)[right=0pt] at (b.east){
    \includegraphics[width=100pt]{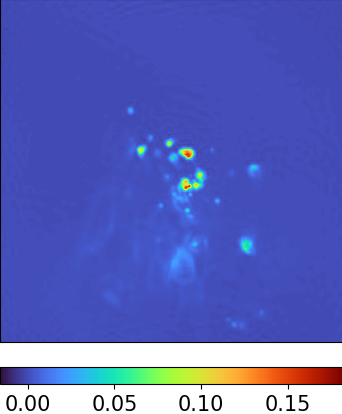}
  };
  \node(cm)[right=0pt] at (c.east){
    \includegraphics[width=100pt]{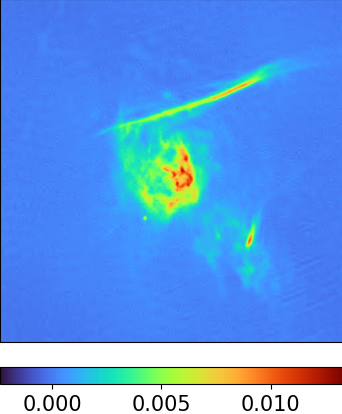}
  };
  
  \node(sa)[text centered, below=5pt, text=white, font=\footnotesize] at (a.north){PSNR (dB) = 17.9};
  \node(sb)[text centered, below=5pt, text=white, font=\footnotesize] at (b.north){PSNR (dB) = 20.8};
  \node(sc)[text centered, below=5pt, text=white, font=\footnotesize] at (c.north){PSNR (dB) = 15.5};
  \node(ma)[text centered, below=5pt, text=white, font=\footnotesize] at (am.north){PSNR (dB) = 21.5};
  \node(mb)[text centered, below=5pt, text=white, font=\footnotesize] at (bm.north){PSNR (dB) = 21.3};
  \node(mc)[text centered, below=5pt, text=white, font=\footnotesize] at (cm.north){PSNR (dB) = 19.1};

  \node(sa_title)[text centered, left=0pt, font=\footnotesize] at (a.west){Sgr A};

  \node(sb_title)[text centered, left=0pt, font=\footnotesize] at (b.west){Sgr B2};

  \node(sc_title)[text centered, left=0pt, font=\footnotesize] at (c.west){Sgr C};

  \node(single)[text centered, above=0pt, font=\large] at (a.north){Single-step};

  \node(multi)[text centered, above=0pt, font=\large] at (am.north){Multi-step};
\end{tikzpicture}
\caption{Final reconstructed images for both the single-step and our proposed multi-step approach. We found our multi-step approach to have slightly better PSNRs than the single-step method.}
\label{Fig:FullPSNR}
\end{figure}
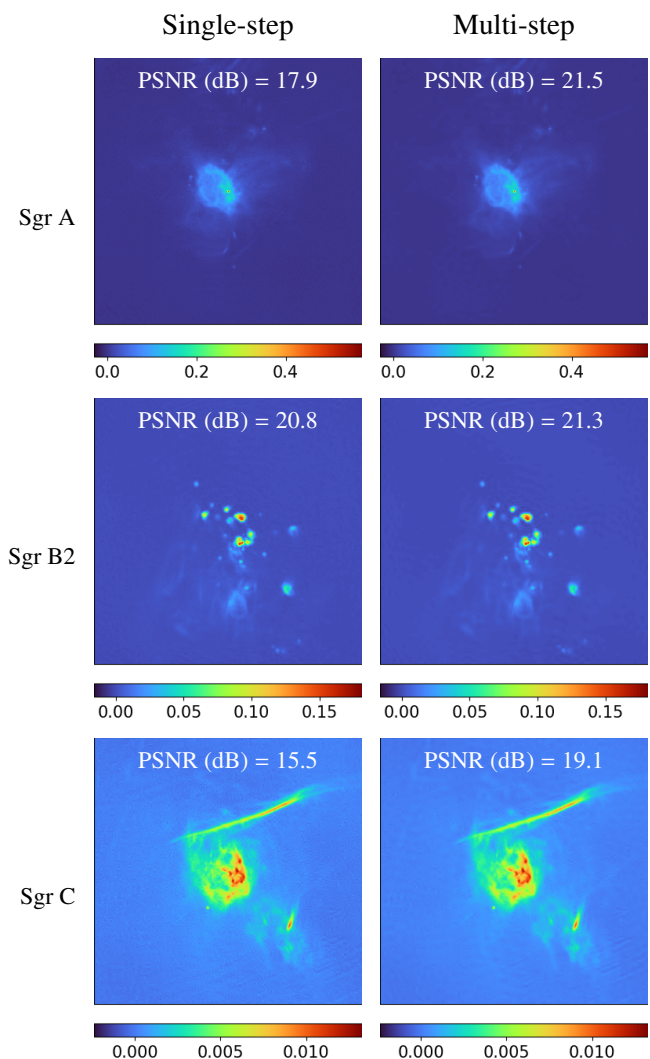

Figure~\ref{Fig:FullPSNR} shows the final reconstructed images with their respective PSNRs in dB for both a single-step reconstruction using all the visibilities $\text{V}_\mathcal{L}\cup \text{V}_\mathcal{H}$ and the proposed multi-step reconstructions using first $\text{V}_\mathcal{L}$ and then $\text{V}_\mathcal{H}$. We found the PSNRs of our multi-step approach to be better than those obtained using the single-step method. This is surprising, as we expected our multi-step approach to introduce additional error between the two steps, resulting in a lower quality image.

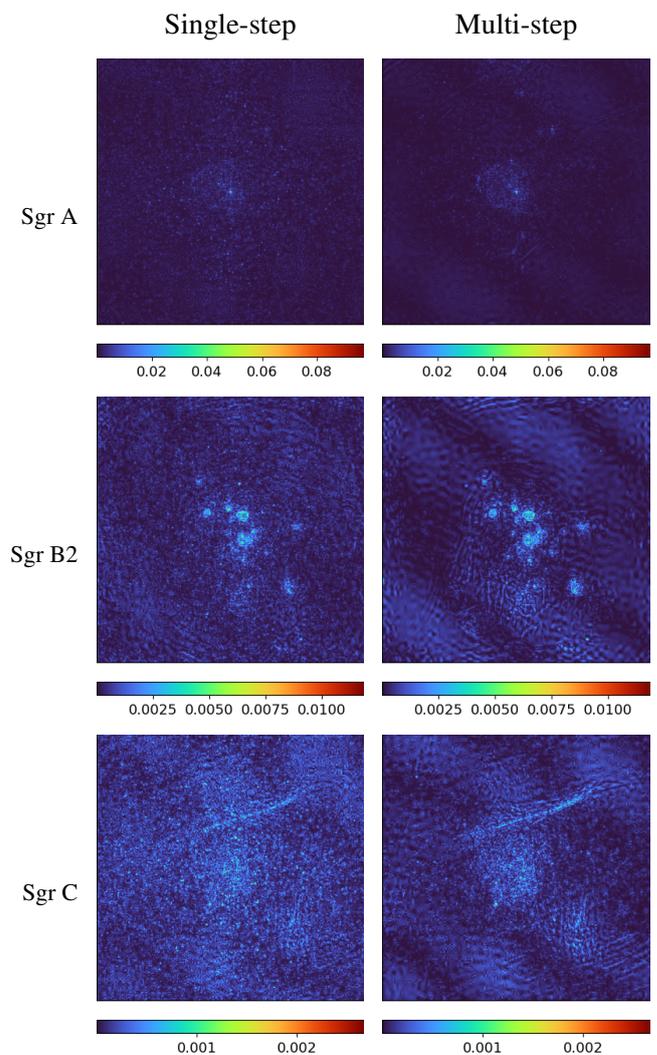
\begin{figure}[t]
\centering
\begin{tikzpicture}
  \node(ae) at (0, 0){
    \includegraphics[width=100pt]{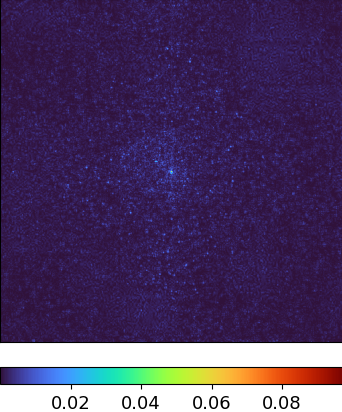}
  };
  \node(be)[below=0pt] at (ae.south){
    \includegraphics[width=100pt]{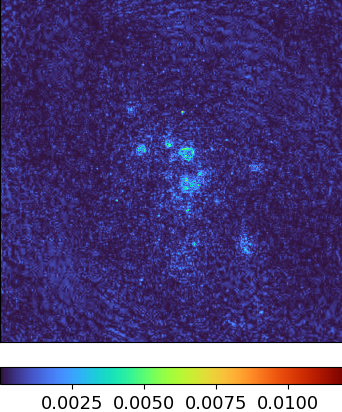}
  };
  \node(ce)[below=0pt] at (be.south){
    \includegraphics[width=100pt]{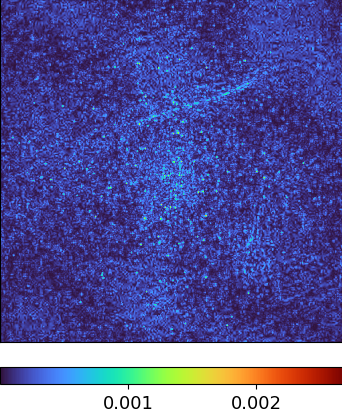}
  };
  
  \node(ame)[right=0pt] at (ae.east){
    \includegraphics[width=100pt]{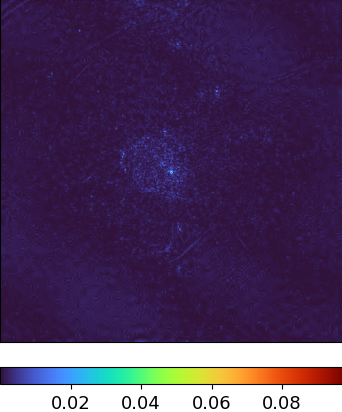}
  };
  \node(bme)[right=0pt] at (be.east){
    \includegraphics[width=100pt]{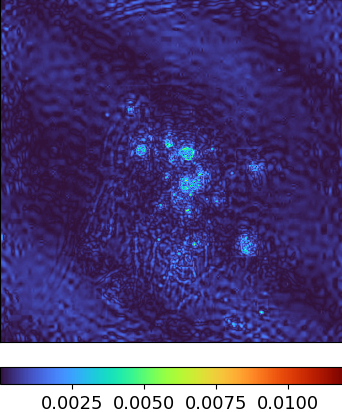}
  };
  \node(cme)[right=0pt] at (ce.east){
    \includegraphics[width=100pt]{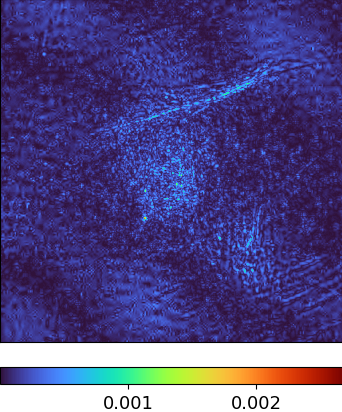}
  };

  \node(sa_title)[text centered, left=0pt, font=\footnotesize] at (ae.west){Sgr A};

  \node(sb_title)[text centered, left=0pt, font=\footnotesize] at (be.west){Sgr B2};

  \node(sc_title)[text centered, left=0pt, font=\footnotesize] at (ce.west){Sgr C};

  \node(single)[text centered, above=0pt, font=\large] at (ae.north){Single-step};

  \node(multi)[text centered, above=0pt, font=\large] at (ame.north){Multi-step};
\end{tikzpicture}
\caption{Absolute error images of the final reconstructed images for both the single and multi-step methods. It can be seen that there is less noise and small-scale error in the multi-step images compared to the single-step.}
\label{Fig:FullErr}
\end{figure}

The main explanation for the better PSNR is that the single-step method has more noise and fine-scale artefacts, which are not as pronounced in the multi-step method, as can be seen in Fig.~\ref{Fig:FullErr}. These are not counterbalanced by the marginally better feature reconstruction, resulting in a lower PSNR for the reconstructed image produced by the single-step method. We hypothesize that this is due to the second data fidelity term of the full-resolution reconstruction step in our multi-step method (Equation~\ref{Eq:FullRes}), which biases the reconstruction towards $\hat{\imath}_\mathcal{L}$. This acts as an additional denoiser, as $\hat{\imath}_\mathcal{L}$ is smoother and less noisy due to both its low-resolution nature and there being more visibilities per grid cell. However, further investigation is required to see if this result is specific to our datasets, or if it can be generalized.

\begin{figure}[t]
\centering
\begin{tikzpicture}
  \node(low) at (0, 0){
    \includegraphics[width=200pt]{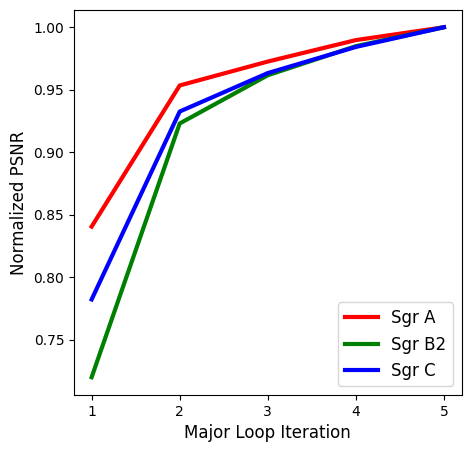}
  };
  \node(high)[below=20pt] at (low.south){
    \includegraphics[width=200pt]{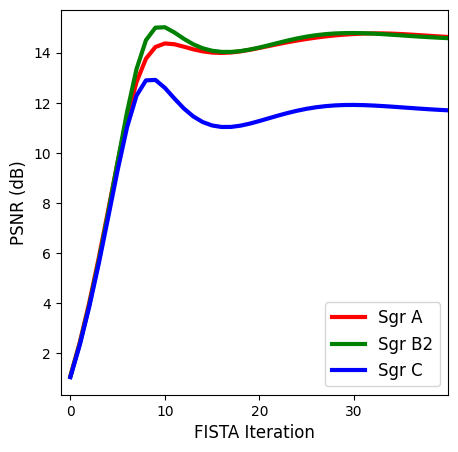}
  };
\end{tikzpicture}
\caption{Progression of reconstruction accuracy of the single-step reconstruction across both major-cycles (top) and FISTA iterations (bottom). A normalized PSNR is used for the top diagram for better comparison of convergence, whereas the actual PSNR values in dB are used for the bottom.}
\label{Fig:FullConvergence}
\end{figure}

We also found the total number of FISTA iterations required by the single-step method to be comparable to the multi-step one. Figure~\ref{Fig:FullConvergence} shows the reconstruction speed across both major-cycles and FISTA iterations. It can be seen that the total number of major cycles required is only slightly more for the multi-step method compared to the single-step one, with around 6-7 required for the former, and 5 or more for the latter.

We also see that in terms of FISTA iterations, the single-step method requires $2-3 \times$ less compared to the full-resolution step in the multi-step method, and requires roughly $1.5 \times$ more than the low-resolution step. This, coupled with there being more major-cycle iterations needed for the low-resolution step than the full, means that the total number of FISTA iterations between the single and multi-step approaches roughly equalize.

\section{Conclusion and Future Work}
\label{S:ConclusionFuturework}
\subsection{Conclusions}
This paper proposes a radio-inteferometric imaging method that allows for partitioning visibilities by baseline. This has several advantages over previous approaches that require all baselines to be treated simultaneously in that:
\begin{itemize}
\item It alleviates memory concerns, as not all baselines need to be gridded simultaneously;
\item It allows more flexible distribution of visibility data in a cluster.
\end{itemize}

We present our method in the context of sparsity regularized convex optimization problems with over-redundant wavelet dictionaries, and compare it to a single-step approach in the same framework that processes all baselines simultaneously without partitioning. In addition to better data distribution, we found our method to produce in our case images with lower amounts of noise and fine-scale artefacts compared to the single-step approach, while not reconstructing perceivably less structure. We also found our method to have comparable computational costs, requiring roughly the same number of total FISTA iterations as the single-step method.

The main drawback to our proposed method is that it creates a secondary data product, the low-resolution image $\hat{\imath}_\mathcal{L}$ used in the full-resolution reconstruction step. This could be a significant stumbling block, especially in the context of the SKA where storage costs are a concern. However, this can be alleviated by reconstructing in 
Equation \ref{Eq:LowRes} a down-sampled image according to the partition configuration.
This smaller sized image then can be incorporated in the second data fidelity term of \ref{Eq:FullRes} after an appropriate modification of $G_\mathcal{L}$ and the addition of a decimation operator to $W\alpha$.

\subsection{Future Work}
\label{SS:FutureWork}

\begin{figure}[t]
\centering
\begin{tikzpicture}[auto, node distance=2cm, >=latex]
\tikzstyle{block} = [draw, minimum height=0.75cm, minimum width=0.8cm, scale=0.8]
\tikzstyle{wrapper} = [draw, minimum height=0.75cm, minimum width=0.8cm, inner sep=9pt]

\node (low_vis) at (0, 0) [above=10pt]{$\text{V}_\mathcal{L}$};

\node (mc1l) at (low_vis.east) [block, right=25pt] {MC};
\node (mc2l) at (mc1l.east) [block, right=20pt] {MC};
\node (mc3l) at (mc2l.east) [block, right=20pt] {MC};
\node (low_ell) at (mc3l.east) [minimum height=0.75cm, minimum width=0.75cm, right=1pt] {$\cdots$};
\node (mcnl) at (low_ell.east) [block, right=0pt] {MC};

\draw [->] (low_vis) edge (mc1l);
\draw [->] (mc1l) -- node [midway, above=0pt] {$i_{\mathcal{L}_1}$} (mc2l); 
\draw [->] (mc2l) -- node [midway, above=0pt] {$i_{{\text{V}_\mathcal{L}}_2}$} (mc3l); 

\node[fit=(mc1l)(mc2l)(mcnl), wrapper](vlnode){};

\node (full_vis) at (low_vis.south)[below=60pt]{$\text{V}_\mathcal{H}$};

\node (mc1h) at (full_vis.east) [block, right=25pt] {MC};
\node (mc2h) at (mc1h.east) [block, right=20pt] {MC};
\node (mc3h) at (mc2h.east) [block, right=20pt] {MC};
\node (low_ell) at (mc3h.east) [minimum height=0.75cm, minimum width=0.75cm, right=1pt] {$\cdots$};
\node (mcnh) at (low_ell.east) [block, right=0pt] {MC};

\draw [->] (full_vis) edge (mc1h);
\draw [->] (mc1h) -- node [midway, above=0pt] {$i_{\mathcal{H}_1}$} (mc2h); 
\draw [->] (mc2h) -- node [midway, above=0pt] {$i_{{\text{V}_\mathcal{H}}_2}$} (mc3h); 

\draw [->] (mc1h) -- node [midway, below left=6pt] {$i_{\mathcal{H}_1}$} (mc2l); 
\draw [->] (mc2h) -- node [midway, below left=6pt] {$i_{{\text{V}_\mathcal{H}}_2}$} (mc3l); 
\draw [->] (mc1l) -- node [midway, above left=6pt] {$i_{\mathcal{L}_1}$} (mc2h); 
\draw [->] (mc2l) -- node [midway, above left=6pt] {$i_{{\text{V}_\mathcal{L}}_2}$} (mc3h); 

\node[fit=(mc1h)(mc2h)(mcnh), wrapper](vhnode){};

\node[fit=(mcnl)(mcnh)](centeringnode){};
\node (comb) at (centeringnode.east)[right=30pt]{$\oplus$};
\node (dummyout) at (comb.east)[right=20pt]{};

\draw [->] (vlnode.east) -- node [midway, right=3pt] {$\hat{\imath}_{\text{V}_\mathcal{L}}$} (comb); 
\draw [->] (vhnode.east) -- node [midway, right=3pt] {$\hat{\imath}_{\text{V}_\mathcal{H}}$} (comb); 
\draw [->] (comb) -- node [midway, above=2pt] {$\hat{\imath}$} (dummyout); 

\end{tikzpicture}
\caption{An example workflow of how parallelism will work with our proposed method. We look to reconstruct to full-resolution images in parallel, achieved by sharing the deconvolved residuals across different nodes after each major cycle, denoted as MC in the figure. A low and high-resolution image are produced after the first major cycle, denoted as $i_{\mathcal{L}_1}$ and $i_{\mathcal{H}_1}$ respectively. Afterwards, full-resolution images $i_{{\text{V}_\mathcal{L}}_n}$ and $i_{{\text{V}_\mathcal{H}}_n}$ are produced. The final reconstructed images of their respective nodes, $\hat{\imath}_{\text{V}_\mathcal{L}}$ and $\hat{\imath}_{\text{V}_\mathcal{H}}$, are then combined to produce the final reconstructed image $\hat{\imath}$.}
\label{Fig:InterleavedMultiStepPipeline}
\end{figure}
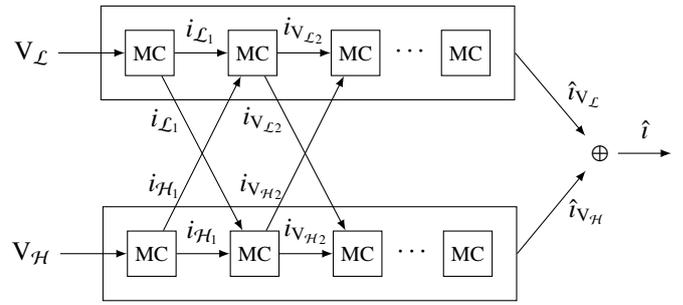

There are several avenues to extending our work. One natural extension is studying how our framework performs with more than 2-steps, which will afford us information on the limit of partitioning by baseline. Another is incorporating parallelism. This can be done by reconstructing two full-resolution images simultaneously, rather than first a low-resolution and then a full, achieved by sharing the deconvolved residual between nodes after every major cycle. Thus, a low and high-resolution image is reconstructed after the first major cycle, and then full-resolution images from there onwards. The full-resolution reconstructions from each node can then be combined to produce the final reconstructed image. An example workflow of this can be seen in Fig.~\ref{Fig:InterleavedMultiStepPipeline}.

Lastly, although we present our method in the framework of sparsity regularized convex optimization problems, the central underlying idea is general. For this reason, an avenue of future research would be to investigate extending this method to other deconvolution frameworks, such as CLEAN or formulations including a positivity constraint.

\begin{acknowledgements}
This work was supported by DARK-ERA (ANR-20-CE46-0001-01)
\end{acknowledgements}

\bibliographystyle{aa}
\bibliography{references}

\begin{appendix}

\section{Preliminary Simulations for Parameter Selection}
\label{App:ParamSelection}
Our proposed method requires knowing the regularization parameters $\lambda_\mathcal{L}$ and $\lambda_\mathcal{H}$, as well as the wavelet dictionary $W$. This appendix discusses our preliminary experiments to determine these. 

\subsection{Choice of Wavelet Dictionary}
\begin{figure}[t]
\centering
\begin{tikzpicture}

  \node(f1) at (0, 0){
    \includegraphics[width=110pt]{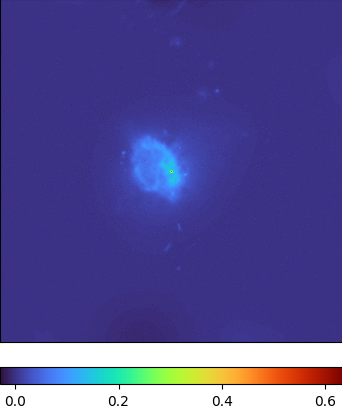}
  };
  \node(f2)[right=1pt] at (f1.east){
    \includegraphics[width=110pt]{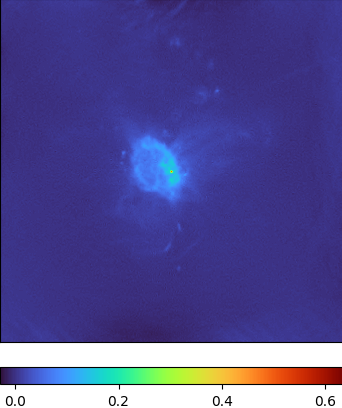}
  };
  \node(f3)[below=1pt] at (f1.south){
    \includegraphics[width=110pt]{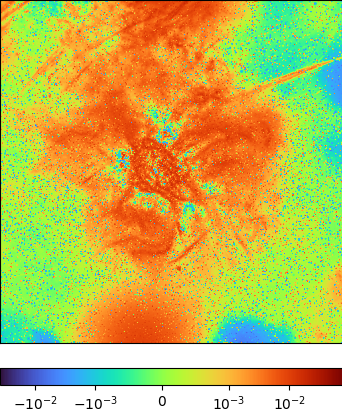}
  };
  \node(f4)[below=1pt] at (f2.south){
    \includegraphics[width=110pt]{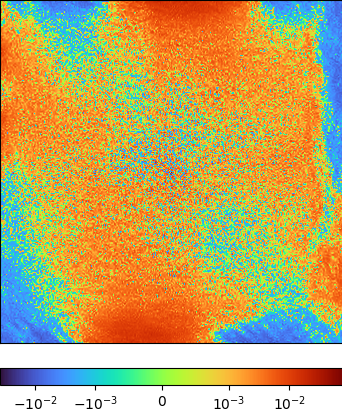}
  };

  \node[text centered, above=0pt, font=\fontsize{3}{0}] at (f1.north){IUWT};
  \node[text centered, above=0pt, font=\fontsize{3}{0}] at (f2.north){Daubechies};

\end{tikzpicture}
\caption{Reconstructed images from the Sgr A full visibilities dataset using both a concatenated dictionary of the first 8 Daubechies wavelets (right) and IUWT (left) wavelets for a single major iteration using a single-step method that reconstructs with $\text{V}_\mathcal{L} \cup \text{V}_\mathcal{H}$. Provided below are the error images against the ground truth, which show that the Daubechies dictionary is able to better reconstruct the large-scale gas structures. We use a $\text{log}_{10}$ scale to better illustrate the structural differences for the error images.}
\label{Fig:wavelets}
\end{figure}

\begin{figure}[t]
\centering
\includegraphics[width=200pt]{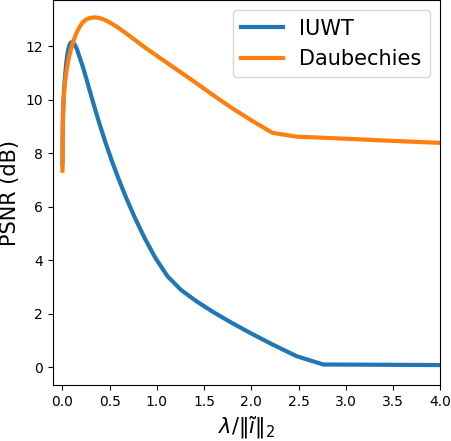}
\caption{PSNR (dB) of different $\lambda$ values for the first major-cycle when performing a single-step reconstruction using $\text{V}_\mathcal{L} \cup \text{V}_\mathcal{H}$ for Sgr A using a concatenation of the first 8 Daubechies wavelets and IUWT wavelets. We found Daubechies wavelets to better reconstruct the large-scale features, as evidenced by the higher PSNR values, and also to be more robust to changes of $\lambda$, having a larger range of values that work well.}
\label{Fig:DaubeIuwtLambda}
\end{figure}

The choice of the wavelet dictionary can greatly affect the quality of the reconstruction. We experimented with using both a concatenation of the first 8 Daubechies~\citep{daubechies1992ten} wavelets and IUWT~\citep{starck2007undecimated} wavelets. Figure~\ref{Fig:wavelets} shows a comparison between reconstructed images of both dictionaries when using a single-step $L^1$ regularized reconstruction of the Sgr A database when simultaneously processing all visibilities $\text{V}_{\mathcal{L} \cup \mathcal{H}}$. 

We found that both dictionaries exhibited artefacts, with Daubechies exhibiting various tiling effects and discontinuities at larger values of $\lambda$, and IUWT exhibiting false sources. However, we found that the Daubechies dictionary allowed for both better reconstruction of large-scale structures, and was more robust to changes in $\lambda$, as shown in Fig.~\ref{Fig:DaubeIuwtLambda}, supporting our use for it in our experiments. A possible reason for why IUWT performs worse for our test cases is due to its isotropic nature, which performs poorly with the large-scale anisotropic features prevalent in our datasets, such as the gas clouds.

\subsection{The regularization parameter \texorpdfstring{$\lambda$}{}}
\label{AS:Lambda}
The selection of the ideal regularization parameters $\lambda_{\mathcal{L}_n}$ and $\lambda_{\mathcal{H}_n}$ depends on a slew of variables, such as the strength and nature of the object being imaged, the amount of noise present in the measurements, and the wavelet dictionary used. We perform preliminary experiments to determine these, with deriving a general strategy being outside the scope of this work. 

All our preliminary experiments are done with the single-step  $L^1$ regularized method that reconstructs with $\text{V}_\mathcal{L} \cup \text{V}_\mathcal{H}$, and evaluate $\lambda$. We generalize the findings here to $\lambda_{\mathcal{L}}$ and $\lambda_{\mathcal{H}}$ for our multi-step method, as the nature of reconstructed images are similar, albeit at different resolutions. Finally, as discussed in the previous section, we use a concatenation of the first 8 Daubechies wavelets as our dictionary.

\begin{figure}[t]
\centering
\begin{tikzpicture}

    \node(f1) at (0, 0){
    \includegraphics[width=110pt]{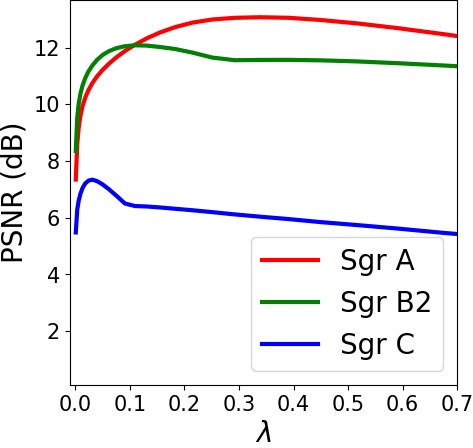}
    };
    \node(f2)[right=0pt] at (f1.east){
    \includegraphics[width=110pt]{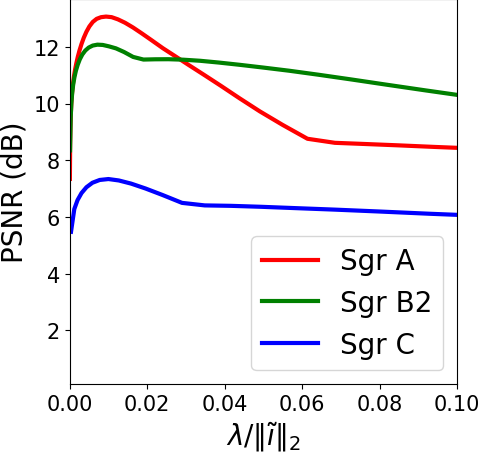}
    };
    
\end{tikzpicture}

\caption{The left image shows PSNR in dB vs $\lambda$ for the first major-cycle for our three initial datasets, illustrating the different ideal values across different objects. This becomes less problematic when normalizing by the $L^2$-norm of the dirty image $\|\tilde{\imath}_1\|_2$. As shown on the right, $\frac{\lambda}{\|\tilde{\imath}_1\|_2}$ is much better behaved, with the region of well performing values all roughly lining up. The reconstruction algorithm used here is the single-step sparsity regularized method that reconstructs using all visibilities without partitioning.}
\label{Fig:FirstMajCycleLambda}
\end{figure}

Figure~\ref{Fig:FirstMajCycleLambda} shows the results for our three datasets for the first major-cycle. The left image plots $\lambda$ as the x-axis, whereas the right plots the normalized value $\frac{\lambda}{\|\tilde{\imath}_1\|_2}$. We found that normalization allows for the well performing regions to line-up, implying that we can pass a constant value for our simulations without the need for dataset specific parameter tuning. We found $\frac{\lambda}{\|\tilde{\imath}_1\|_2} = 0.01$ to work well for the single-step reconstructions, and $\frac{\lambda_{\mathcal{L}}}{\|\tilde{\imath}_{\mathcal{L}_1}\|_2} = \frac{\lambda_{\mathcal{H}}}{\|\tilde{\imath}_{\mathcal{H}_1}\|_2} = 0.05$ to work well for our multi-step method. 

We hypothesize that the larger ideal values for the low-resolution step can be explained by there being less frequencies, ergo needing less wavelet coefficients. On the other hand, the larger values for the full-resolution step could be explained by the low-resolution data fidelity term biasing the reconstruction to have less noise and fine-scale features, which also increases the sparsity of the wavelet coefficients.

\begin{figure}[t]
\centering
\includegraphics[width=200pt]{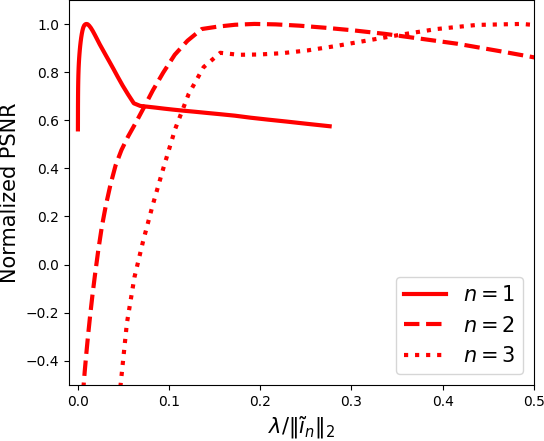}
\caption{Normalized PSNR against $\frac{\lambda}{\|\tilde{\imath}_n\|_2}$ for the first three major cycles when reconstructing the Sgr A database using the single-step method that reconstructs using $\text{V}_\mathcal{L} \cup \text{V}_\mathcal{H}$. The ideal values for $\frac{\lambda}{\|\tilde{\imath}_n\|_2}$ increase as we progress through the major cycles.}
\label{Fig:NMajorCycleLambda}
\end{figure}

\begin{figure}[t]
\centering
\begin{tikzpicture}

    \node(f1) at (0, 0){
    \includegraphics[width=110pt]{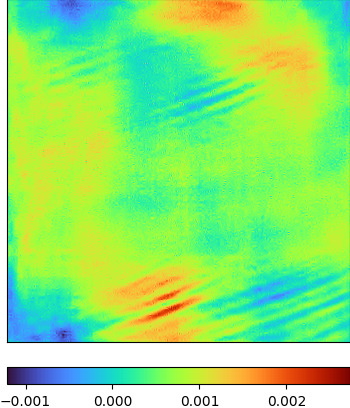}
    };
    \node(f2)[right=0pt] at (f1.east){
    \includegraphics[width=110pt]{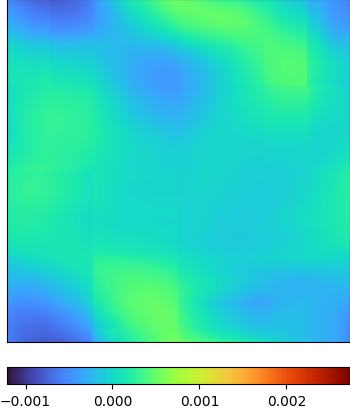}
    };
    \node(f3)[below=0pt] at (f1.south){
    \includegraphics[width=110pt]{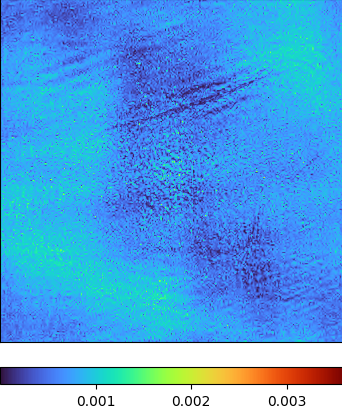}
    };
    \node(f4)[below=0pt] at (f2.south){
    \includegraphics[width=110pt]{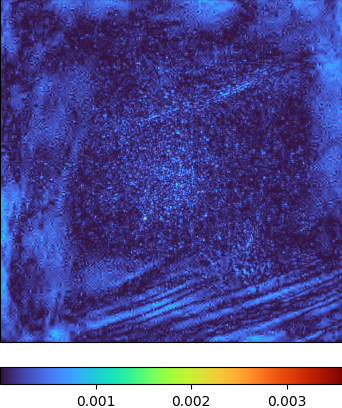}
    };

    \node[text centered, above=0pt, font=\small] at (f1.north){$\lambda=0.018$, PSNR (dB)=-6.74};
    \node[text centered, above=0pt, font=\small] at (f2.north){$\lambda=0.516$, PSNR (dB)=2.73};
    
\end{tikzpicture}
\caption{Reconstructed images as well as their absolute errors of the Sgr C dataset for the second major cycle residual image $\tilde{\imath}_2$ using the single-step method, with a lower (left) and higher (right) $\lambda$. The larger $\lambda$ has better PSNR as it reduces the error on a large scale, but is not able to reconstruct any of the finer-scale details.}
\label{Fig:NMajorCycleLambdaImages}
\end{figure}

We also evaluate how $\lambda$ changes across major-cycle iterations, which is necessary as the nature of residual images $\tilde{\imath}_n$ changes. Figure~\ref{Fig:NMajorCycleLambda} illustrates the behaviour of $\frac{\lambda}{\|\tilde{\imath}_n\|_2}$ across three different major cycles for the Sgr A dataset. We normalize by the maximum PSNR obtained for each respective major-cycle for easier comparison. The ground truth images we use to compute the PSNR are defined as $I - \sum_N \hat{i}_n$, where $I$ is the initial ground truth, and $\hat{i}_n$ is the best reconstructed image amongst the different $\frac{\lambda}{\|\tilde{\imath}_n\|_2}$ values for the $n$th major-cycle.

It can be seen that the ideal values for $\frac{\lambda}{\|\tilde{\imath}_n\|_2}$ generally increases as we progress through the major-cycle iterations. This is expected as the residual images are increasingly dominated by noise, necessitating larger values of $\frac{\lambda}{\|\tilde{\imath}_n\|_2}$ to suppress it. However, care needs to be taken not to increase it too aggressively, as this will lead to all fine scale features being ignored, as shown in Fig.~\ref{Fig:NMajorCycleLambdaImages}. We found that multiplying $\frac{\lambda}{\|\tilde{\imath}_n\|_2}$ by $2^n$ for the $n$th major cycle provides a good trade-off between the two. 
\end{appendix}

\end{document}